\documentclass[conference,a4paper,onecolumn]{IEEEtran}

\usepackage{amsmath, amsthm, amssymb}
\usepackage{amsfonts}
\usepackage{amsthm}
\usepackage{tikz}
\usetikzlibrary{shapes,arrows}
\usepackage{verbatim}
\usepackage[]{subfigure}
\theoremstyle{remark}
\usepackage{multicol}

\usepackage{graphicx,relsize,verbatim,enumerate,xcolor}
\usepackage{pgfplots}

\usetikzlibrary{arrows,positioning,fit,backgrounds,shapes}
\usetikzlibrary{calc}

\newtheorem{lemma}{Lemma}
\newtheorem{theorem}{Theorem}
\newtheorem{definition}{Definition}

\newcommand{\distnarg}[2][p]{\ensuremath{\MakeUppercase{#1}_{#2}}}

\newcommand{\rightratearrow}[5]{
\draw ([yshift=#3] #1) to node [midway,above] {#5} ([yshift=#3,xshift=-#4] #2) -- ++(0,#3) -- (#2);
\draw ([yshift=-#3] #1) -- ([yshift=-#3,xshift=-#4] #2) -- ++(0,-#3) -- (#2);
                            }

\tikzstyle{arw}=[->,>=latex]
\tikzstyle{node}=[rectangle,draw,outer sep=0pt,minimum width=1.7cm, minimum height=8mm]

\newcommand{\clos}{\mbox{Closure}}

\newcommand{\E}{{\mathbb{E}}}
\newcommand{\br}[1]{{\left(#1\right)}}
\newcommand{\setbr}[1]{{\left\{#1\right\}}}
\newcommand{\abs}[1]{{\left|#1\right|}}

\newcommand{\R}{{\mathbb{R}}}

\newcommand{\defeq}{{~\triangleq~}}
\newcommand{\p}{{\mathbb{P}}}

\newcommand{\inv}{{^{-1}}}

\newcommand{\D}{{\Delta}}
\newcommand{\Up}{{\Upsilon}}

\newcommand{\X}{{\mathcal{X}}}
\newcommand{\Y}{{\mathcal{Y}}}
\newcommand{\Z}{{\mathcal{Z}}}
\newcommand{\U}{{\mathcal{U}}}
\newcommand{\V}{{\mathcal{V}}}
\newcommand{\C}{{\mathcal{C}}}
\newcommand{\B}{{\mathcal{B}}}
\newcommand{\s}{{\mathcal{S}}}

\newcommand{\ex}{{\exists}}
\newcommand{\e}{{\epsilon}}
\newcommand{\es}{{\emptyset}}

\newcommand{\setb}[1]{{\left\{#1\right\}}}

\newcommand{\normtv}[1]{{\left\|#1\right\|_{\mbox{\tiny{$TV$}}}}}

\begin{document}

\sloppy

\title{Secure Cascade Channel Synthesis} 

\author{
  \IEEEauthorblockN{Sanket Satpathy and Paul Cuff}
}

\maketitle

\let\thefootnote\relax\footnote{The authors are supported by the National Science Foundation (grants CCF-1116013 and CCF-1350595) and the Air Force Office of Scientific Research (grants FA9550-12-1-0196 and FA9550-15-1-0180).  Portions of this work were presented in \cite{sai}.

The authors are with the Department of Electrical Engineering, Princeton University, Princeton, NJ, 08544 USA (email: satpathy@princeton.edu; cuff@princeton.edu)}
\begin{abstract}
  We consider the problem of generating correlated random variables in a distributed fashion, where communication is constrained to a cascade network.  The first node in the cascade observes an i.i.d.\ sequence $X^n$ locally before initiating communication along the cascade. All nodes share bits of common randomness that are independent of $X^n$. We consider secure synthesis - random variables produced by the system appear to be appropriately correlated and i.i.d.\ even to an eavesdropper who is cognizant of the communication transmissions. We characterize the optimal tradeoff between the amount of common randomness used and the required rates of communication. We find that not only does common randomness help, its usage exceeds the communication rate requirements.  The most efficient scheme is based on a superposition codebook, with the first node selecting messages for all downstream nodes. We also provide a fleeting view of related problems, demonstrating how the optimal rate region may shrink or expand.
\end{abstract}

\section{Introduction}


This paper studies the synthesis of correlated random variables under a total variation constraint, known as strong coordination \cite{coord} or channel synthesis \cite{DCS}.  Given an i.i.d.\ input to the system, we would like to generate a correlated output at a remote location - this can be understood as approximation of a conditional probability distribution. Due to the stochastic nature of the objective, the cooperating parties benefit from access to common randomness, in addition to their communication capabilities.

The optimal tradeoff between communication and common randomness was derived by Cuff \cite{Cuff1} and Bennett et al. \cite{reverse2} for the case of two random variables and the results have been extended in other work \cite{Gohari1,Gohari2,Gohari3,Gohari4,DCS}. One particularly pleasing aspect of the above tradeoff was that it recovered two familiar measures of correlation as the required rate of communication --- mutual information and Wyner's common information \cite{Wyner} --- in the presence of abundant and no common randomness, respectively.

Requiring that the synthesized joint distribution be close to the desired joint distribution in total variation is a more stringent constraint than empirical coordination \cite{Haim,coord} i.e.\ jointly typical input and output sequences. On the other hand, we enjoy the benefit of the synthesized sequences being immune to statistical tests designed to detect i.i.d.\ correlated sequences \cite{DCS}.  This is a simple consequence of the fact that hypothesis tests will produce identically distributed outcomes for distributions that are extremely close in total variation.  Also, total variation can be bounded by entropy \cite[Theorem 17.3.3]{Cover}, which is extremely useful for proving converse results.

The above observations have led to applications in secrecy and game theory \cite{Cuff1,Cuff2}.  Schieler and Cuff \cite{RDSS} study secrecy with causal disclosure of information to the eavesdropper - their achievability scheme hinges on a distributional approximation result.  Winter \cite{Winter05} and Chitambar et al.\ \cite{Winter14} consider secure generation of correlated random variables along the lines of the channel synthesis problem \cite{DCS}, with the notable difference that no information sequence is provided as an external input.  We refer the interested reader to  \cite{DCS,Bloch2} for a more thorough discussion of the properties of total variation and comparison with other metrics.

\tikzstyle{block} = [draw, fill=blue!20, rectangle, 
    minimum height=2em, minimum width=4em]
\tikzstyle{sum} = [draw, fill=blue!20, circle, node distance=1cm]
\tikzstyle{input} = [coordinate]
\tikzstyle{output} = [coordinate]
\tikzstyle{pinstyle} = [pin edge={to-,thin,black}]

\begin{figure}[ht]
\begin{center}
\resizebox{3.6in}{1.5in}{
\begin{tikzpicture}[scale=1,auto, node distance=1cm,>=latex']
    \node [input, name=input] {};
    \node [block, below of=input, node distance=1.5cm] (node1) {$F_n$};
    \node [block, right of=node1,
            node distance=4cm] (node2) {$G_n$};
    \node [block, right of=node2,
            node distance=4cm] (node3) {$H_n$};
    \draw [->,double] (node2) -- node[name=r2] {$nR_2$ bits} (node3);
    \draw [] (node3) -- node[name=j2] {$M_2$} (node2);
    \node [output, below of=node2, node distance=1.5cm] (output1) {};
    \node [output, below of=node3, node distance=1.5cm] (output2) {};
    \node [draw=none, above of=node2, node distance = 2cm] (commrand) {$nR_0$ bits};

    \draw [draw,<-] (node1) -- node {$X^n$} (input);
    \draw [->,double] (node1) -- node {$nR_1$ bits} (node2);
    \draw [] (node2) -- node {$M_1$} (node1);
    \draw [->] (node2) -- node [name=yn] {$Y^n$}(output1);
    \draw [->] (node3) -- node [name=zn] {$Z^n$}(output2);
    \draw [->,double] (commrand) -- (node1);
    \draw [->,double] (commrand) -- (node2);
    \draw [->,double] (commrand) -- (node3);
\end{tikzpicture}}
\caption{The i.i.d.\ sequence $X^n$ is given by nature. Messages $M_1$ and $M_2$ are sent along the cascade at rates $R_1$ and $R_2$. Common randomness $K$ is shared by all 3 nodes at rate $R_0$. We want $(X^n,Y^n,Z^n)$ i.i.d.\ correlated and independent of the messages $(M_1,M_2)$.}\label{sccs}
\end{center}
\end{figure}
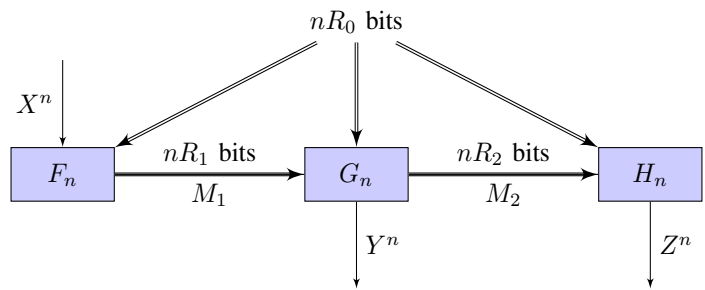

This work extends \cite{DCS} to a network setting. Such extensions have been previously considered by \cite{Gohari1,Gohari4}, albeit without tight results. The cascade channel synthesis problem shown in Fig. \ref{sccs} provides a starting point for the theory of strong coordination over general networks. Distributed networks for control and sensing are seemingly ubiquitous - the power grid, road networks, server farms and the internet are a few prominent examples \cite{control}. In many control settings, one would like the actions at various nodes to be independent of the communication so the actions cannot be anticipated by malicious eavesdroppers. We consider the problem of synthesizing sequences that appear to be i.i.d.\ and appropriately correlated even from the perspective of an onlooker who can see messages sent over the communication channel but does not have access to the source of common randomness.

The cascade structure for communication is especially relevant for large distributed networks where there is a cost per unit distance associated with sending messages. In such a setting, it would be more economical for nodes to forward appropriate messages locally in a cascade fashion rather than having a central node talk to all the other nodes. Also in some settings, there might be a hierarchy among the nodes that inherits the cascade structure (for example, in parallel computation, interaction takes place in a master-slave hierarchy). In section II-E, we characterize the optimal rate region for the task assignment problem, where 3 out of $m\ge3$ tasks have to be distributed uniformly at random to the 3 nodes, with the first node's tasks chosen at random by nature.  In section III-A, this example is used to demonstrate how the rate region is sensitive to the particulars of the secrecy constraint.

In the context of source coding, the cascade network has received some prior attention.  Yamamoto \cite{casca1} considered the lossy transmission of a source pair $(X,Y)$ to receivers along a cascade, with $X$ and $Y$ intended for separate receivers.  Vasudevan, Tian and Diggavi \cite{casca2} consider the case when $X=Y$ i.e.\ both receivers desire the same data, and they have additional side-information.  Permuter and Weissman \cite{casca4} also explore the role of side-information at the initial nodes and provide an explicit solution for Gaussian random variables.  Cuff, Su and El Gamal \cite{casca3} study a variant where $X$ and $Y$ are provided at the first and second nodes of the cascade, while the receiving node seeks a function $f(X,Y)$.

All of the above works can be broadly categorized as problems of empirical coordination \cite[Theorem 11]{coord}.  In contrast, we study strong coordination \cite{coord} in the cascade network.  Cuff, Permuter and Cover \cite[Theorem 5]{coord} solve the analogous problem for empirical coordination on arbitrarily long cascade networks.  In the same paper, they make a conjecture about the relationship between optimal schemes for empirical coordination and strong coordination.  In section III-B, we provide an example that refutes the most general version of their conjecture.

In this work, we only consider random variables with finite alphabets and assume that every node in the cascade has sufficient local randomness. Interested readers may refer to \cite{DCS} for a treatment of synthesis with limited local randomness.  Bloch and Kliewer \cite{Bloch3} studied the cascade synthesis problem without secrecy constraints.  They provide inner and outer bounds that do not match in general.  In fact, we find that the secrecy constraint makes a complete characterization of the rate region tractable.  Vellambi et al. \cite{Vell, Vell1} also study the tradeoff w.r.t.\ local randomness at the nodes without secrecy constraints.  They provide a general achievability scheme that is proved to be optimal in certain regimes of common randomness, with additional assumptions on the encoder, or when the distribution to be synthesized enjoys a Markov chain structure.  The functional regime considered in \cite{Vell1} coincides with optimal operation in the secure cascade synthesis problem.  Finally, \cite{Cuff6} uses ideas from the present work to devise secure source coding schemes for a cascade network.

\section{Main Result}

\subsection{Notation}

We represent both random variables and probability distribution functions with capital letters, but only letters $P$ and $Q$ are used for the latter. The conditional distribution of the random variable $Y$ given the random variable $X$ is denoted by $P_{Y|X}(y|x)$. We may abbreviate this as $P_{Y|X}$. We also abbreviate the product distribution $\prod_{i=1}^nP_{X|Y}(x_i|y_i)$ as $\prod P_{X|Y}$ when the $y^n$ sequence is clear from context. We use the script letter $\X\ni x$ to denote the alphabet of $X$. Sequences $X_i,\ldots,X_n$ are denoted by $X_i^n$, with $X^n:=X_1^n$. The set $\setb{1,\ldots,m}$ is denoted by $[m]$.

Markov chains are denoted by $X-Y-Z$ implying the factorization $P_{XYZ}=P_{XY}P_{Z|Y}$. The factorization $P_{XY}=P_XP_Y$, i.e.\ independence, is denoted by $X\perp Y$. We define the total variation distance as
\begin{equation}
\normtv{P_X-Q_X}\defeq\frac{1}{2}\sum_x\abs{P_X(x)-Q_X(x)}.\label{tv}
\end{equation}
The convex hull of a set $A$ is denoted by $\mbox{Conv}(A)$.
The empirical distribution (normalized counts) of a vector $x^n$ is denoted by $\p_{x^n}\in\D^{\abs{\X}-1}$, where
\begin{equation}
\p_{x^n}(y)=\frac{1}{n}\sum_{i=1}^n1_\setbr{x_i=y}
\end{equation}
for $y\in\X$, and $\D^k$ denotes the $k$-dimensional simplex.  Wyner's common information \cite{Wyner} between random variables $X$ and $Y$ is denoted by $C(X;Y)$, and defined as
\begin{equation}
C(X;Y)=\min_{U:X-U-Y}I(X,Y;U).
\end{equation}

\subsection{Problem-Specific Definitions}

Although we solve the synthesis problem for arbitrarily long cascades, we restrict the presentation here to a cascade of three nodes for simplicity.  We state our most general result in section II-F.

We have the i.i.d.\ source $X^n\sim\prod_{i=1}^nQ_X$ and we would like to \emph{synthesize} the channel $\prod Q_{YZ|X}$. Messages sent along the cascade communication links are denoted by $M_1\in[2^{nR_1}]$ and $M_2\in[2^{nR_2}]$. The common randomness shared by all nodes $K$ is uniformly distributed on $[2^{nR_0}]$ and independent of $X^n$.

\begin{definition}
A $(2^{nR_0},2^{nR_1},2^{nR_2},n)$ \emph{secure cascade channel synthesis (SCCS) code} consists of randomized encoding functions
\begin{IEEEeqnarray*}{rCl}
F_n&:&\X^n\times[2^{nR_0}]\to[2^{nR_1}],\\
G_n^{(enc)}&:&[2^{nR_1}]\times[2^{nR_0}]\to[2^{nR_2}],\vspace{-0.1cm}
\end{IEEEeqnarray*}
and randomized decoding functions
\begin{IEEEeqnarray*}{rCl}
G_n^{(dec)}&:&[2^{nR_1}]\times[2^{nR_2}]\times[2^{nR_0}]\to\Y^n,\\
H_n&:&[2^{nR_2}]\times[2^{nR_0}]\to\Z^n.
\end{IEEEeqnarray*}
The randomization used for each of these functions is assumed to be independent.  In other words, a randomized function can be thought of as a function with an additional argument that is a random variable (suppressed in the notation).  Each function uses a random variable which is independent of everything else in the problem setting.
We have $M_1=F_n(X^n,K),M_2=G_n^{(enc)}(M_1,K)$, $Y^n=G_n^{(dec)}(M_1,M_2,K)$ and $Z^n=H_n(M_2,K)$.
\end{definition}

Note that node 2 has both encoding and decoding capability. The \emph{induced joint distribution} of a $(2^{nR_0},2^{nR_1},2^{nR_2},n)$ SCCS code is the joint distribution $P_{X^n,Y^n,Z^n,K,M_1,M_2}$ as per the above specifications.

\begin{definition}
A sequence of $(2^{nR_0},2^{nR_1},2^{nR_2},n)$ SCCS codes for $n\ge1$ is said to \emph{achieve} $Q_{YZ|X}$ if the induced joint distributions have marginals that satisfy
\begin{equation}
\lim_{n\to\infty}\normtv{P_{X^nY^nZ^nM_1M_2}-P_{M_1M_2}\prod_{t=1}^nQ_{XYZ}}=0.\label{synthdef}
\end{equation}
\end{definition}

\begin{definition}
A rate triple $(R_0,R_1,R_2)$ is said to be \emph{achievable} if there exists a sequence of $(2^{nR_0},2^{nR_1},2^{nR_2},n)$ SCCS codes that achieves $Q_{YZ|X}$.
\end{definition}

\begin{definition}
The \emph{secure synthesis rate region} $\mathcal{C}$ is the closure of achievable triples $(R_0,R_1,R_2)$.
\end{definition}

\subsection{Main Result}
The characterization of the set of achievable rate triples is given in terms of the following set
\begin{equation}\s_{D}\defeq \setbr{\begin{array}{r c l}
\hspace{-.1cm}(R_0,R_1,R_2)\in\R^3&:&\ex P_{X,Y,Z,U,V}\in D \mbox{ s.t.}\\
R_1&\ge&I(X;U,V)\\
R_2&\ge&I(X;V)\\
R_0&\ge&I(X,Y,Z;U,V)
\end{array}},\label{result}\end{equation}where
\begin{equation}D\defeq\setbr{\begin{array}{r c l}
P_{X,Y,Z,U,V}&:&(X,Y,Z)\sim Q_XQ_{YZ|X},\\
&&X-(U,V)-Y,\\
&&(X,Y,U)-V-Z,\\
&&\abs{\V}\le\abs{\X}\abs{\Y}\abs{\Z}+3,\\
&&\abs{\U}\le\abs{\X}\abs{\Y}\abs{\Z}\abs{\V}+3
\end{array}}.\label{D}\end{equation}Also, let $D'=D\cap\setbr{P_{X,Y,Z,U,V}:H(V|U)=0}$ denote the restriction of $D$ to joint distributions where $V$ is a function of $U$.

\begin{theorem}
\begin{equation}\C=\s_D=\s_{D'}.\end{equation}\label{thm}\end{theorem}
\vspace{-0.6cm}

The achievability proof is based on superposition coding - the first node picks messages for all links and downstream nodes merely forward the appropriate messages. The details are presented in section IV.

\subsection{Remarks}

A startling feature of the optimal encoding scheme, which is implied by the optimal region above,  is that there is no loss of generality in assuming that the second message is a function of the first i.e.\ the local randomness used in synthesizing $Y^n$ is not essential to correlating $Z^n$ with $(X^n,Y^n)$. However, it is precisely this feature that obstructs solution to the cascade channel synthesis problem with no eavesdropper \cite{Bloch3}.  In the absence of secrecy constraints, the system may benefit when intermediate nodes generate messages locally - easily seen when $X\perp(Y,Z)$ i.e.\ the first node is passive.

For the above case, our result reduces to the region $R_0\ge C(Y;Z)$, obtained by picking $U=\es$ and $V\perp X$, with $Y-V-Z$ to minimize $I(V;Y,Z)$. Communication is not necessary, or useful in any way, even between the active nodes! In the non-secure synthesis problem, the optimal rate region is given by $R_2+R_0\ge C(Y;Z)$.  We see that communication (originating at the second node) is still not necessary but can be useful.  Finally, consider a modified secure synthesis problem, where the $Y^n$ sequence is provided by nature (with $X$ still independent of $(Y,Z)$).  In this case, the optimal rate region \cite[III.C]{DCS} is described by the constraints
\begin{align}
R_2&\ge I(Y;V),\\
R_0&\ge I(Y,Z;V),
\end{align}
with $Y-V-Z$.  Communication is necessitated by external inputs to the system.

In the regime of abundant common randomness, the communication rate requirements of our result coincide with the rate region for empirical coordination in the cascade channel \cite[Theorem 5]{coord} since there must exist a realization of the shared randomness that yields good empirical coordination codes, in agreement with \cite[Theorem 2]{coord}. This observation consolidates the intuition that the onus of secrecy that we have taken on is borne primarily by the available common randomness.  However, note that the rate of common randomness can be much larger than the largest communication rate.  This shows again that while common randomness is helpful for secrecy, it also serves to coordinate the actions of the nodes.


By the data-processing inequality \cite{Cover}, the choice $(U,V)=(Y,Z)$ simultaneously minimizes both the communication rates at $R_1\ge I(X;Y,Z)$ and $R_2\ge I(X;Z)$. Also, the minimum achievable rate of common randomness is
\begin{equation}C_c(X;Y;Z)=\min_{(U,V):\substack{X-(U,V)-Y,\\(X,Y,U)-V-Z}}I(X,Y,Z;U,V),\label{casc}
\end{equation}
which can be viewed as a generalization of Wyner's common information in the cascade setting. Another straightforward generalization of Wyner's common information that has been considered in the literature \cite{comm} is\vspace{-0.1cm}
\begin{equation}C(X;Y;Z)=\min I(X,Y,Z;U),\label{wyn}
\end{equation}
where the minimum is over random variables $U$ such that given $U$, $X,Y$ and $Z$ are independent of each other. In fact, the two quantities are the same (without regard to constraints on the alphabet sizes).
This is because the minimizers of \eqref{wyn} and \eqref{casc} are compatible with each other's Markov structures. For example, if $\hat{U}$ attains the minimum for \eqref{wyn}, then $(U,V)=(\es,\hat{U})$ satisfies the Markov chains in \eqref{casc}. Note that the communication and common randomness rates cannot be simultaneously minimized in general.\vspace{-0.1cm}

\subsection{Task Assignment Example}

We now compute our region for an example we will call \emph{task assignment}, where 3 out of $m\ge3$ tasks are to be assigned to the 3 nodes uniformly at random. To be precise, we consider a channel $Q_{YZ|X}$ that acts on $X$ uniformly distributed on $[m]$ i.e. $Q_X=m\inv$ and produces a pair $Y\neq Z$ uniformly distributed over all distinct pairs in $[m]\setminus\setbr{X}$. This is a generalization of the scatter channel example in  \cite{DCS}.

We shall now explain how the Markov chains in \eqref{result} simplify the computation of the rate region by letting us parametrize $(U,V)$.  As per \eqref{result}, we consider joint distributions $P_{X,Y,Z,U,V}\in D$. The Markov chains ensure that the conditional distribution factorizes as $P_{X,Y,X|U,V} = P_{X|U,V} P_{Y|U,V} P_{Z|V}$.  Also, these distributions have the constraint that the supports of $X,Y$ and $Z$ cannot intersect.

By the above Markov constraints, for any $v \in {\cal V}$, the conditional distribution $P_{X,Y,Z|V=v}$ must have support$(Z)$ disjoint from support$(X)$ $\cup$ support$(Y)$.  Let $a_v = m - |$support$(Z)| \geq |$support$(X)$ $\cup$ support$(Y)|$ for this conditional distribution.  If we further condition on $U = u$, for any $u \in {\cal U}$, then $P_{X,Y,Z|U=u,V=v}$ must have all three supports of $X, Y$ and $Z$ disjoint.  Let $b_{u,v} = |$support$(Y)| < a_v$.  This leaves $|$support$(X)| \leq a_v - b_{u,v}$.  From these observations alone, we can bound the entropies and thus the rate region.
We have the parametric bounds
\begin{IEEEeqnarray}{rCl}
H(X|V=v)&\le&\log a_v\\
H(X|U=u,V=v)&\le&\log(a_v-b_{u,v})\\
H(X,Y,Z|U=v,V=v)&\le&\log(a_v-b_{u,v})b_{u,v}(m-a_v),
\end{IEEEeqnarray}
which do not depend on the marginal distribution $P_{UV}$.  The above inequalities give lower bounds on the required rates. We can achieve these rates by letting $U$ and $V$ be uniformly distributed over all permissible subsets for all choices of $a_v, b_{u,v}$. Thus, the rate region is given by the convex hull of the set

\begin{equation}\setbr{\begin{array}{r c l}
R_0^2\in\R^3&:&\ex a\in[m-1]\setminus\{1\}, b\in[a-1]\mbox{ s.t.}\\
R_1&\ge&\log\br{\frac{m}{a-b}}\\
R_2&\ge&\log\br{\frac{m}{a}}\\
R_0&\ge&\log\br{\frac{m(m-1)(m-2)}{(a-b)b(m-a)}}\\
\end{array}}.\label{task}\end{equation}

\begin{figure}[ht]
\begin{center}
  \pgfimage[height=10cm]{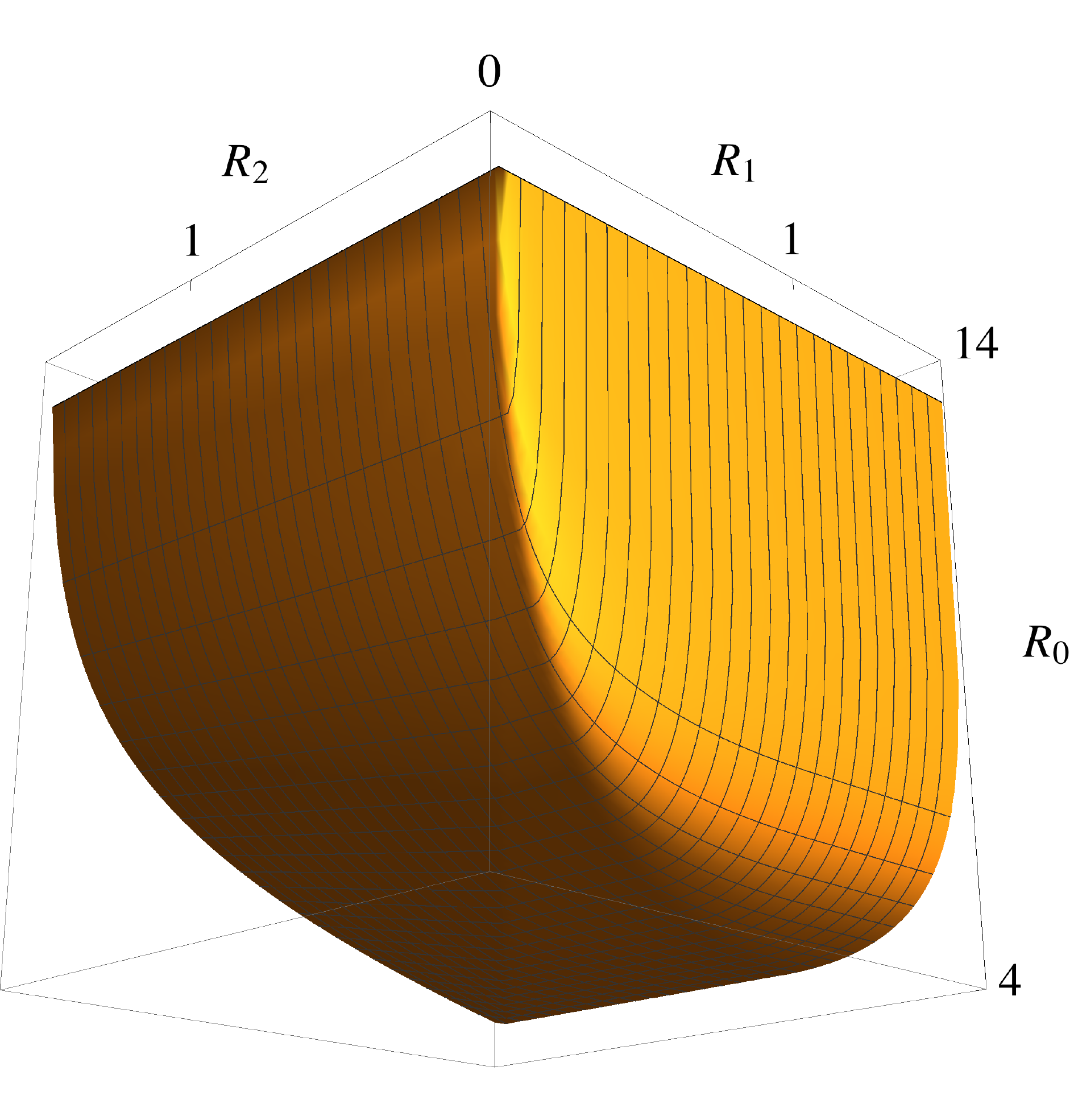}
   \caption{A slice of the rate region \eqref{task} for $m=100$ tasks.  For schemes near the optimal frontier, the common randomness rate $R_0$ far exceeds the communication rates.  Due to the structure of the optimal scheme in Theorem 1, the dominant tradeoff is between $R_1$ and $R_0$.  In general, the optimal rate surface is polyhedral with up to $\frac{(m-1)(m-2)}{2}$ vertices.}
    \label{taskfig}
  \end{center}
  \end{figure}

The communication rates are minimized when $a=m-1$ and $b=1$ i.e.\ given $(U,V)$, the uncertainty is concentrated on $X$. On the other hand, the common randomness requirement is minimized when $b\approx\frac{m}{3}$ and $a\approx\frac{2m}{3}$ up to the nearest integer i.e.\ given $(U,V)$, the uncertainty is shared equally by $X,Y$ and $Z$. The tradeoff between information content of the messages and rate of common randomness is evident  here.  A plot of the optimal rate region is provided in Fig.\ \ref{taskfig}.

\subsection{Arbitrarily Long Cascades}


Our main result of Theorem \ref{thm} can be readily extended to secure channel synthesis of a distribution $Q_{Y_1,\ldots,Y_{m-1}|X}$ for a cascade with $m\ge3$ nodes, with communication rates $R_1,\ldots,R_{m-1}$ on the links of the cascade and common randomness shared by all nodes at rate $R_0$.

The optimal rate region is
\begin{equation} \setbr{\begin{array}{r c l}
R_0^{m-1}\in\R^m&:&\ex P_{X,Y_1^{m-1},U_1^{m-1}}\in D_m \mbox{ s.t.}~\forall~1\le i\le m-1,\\
R_i&\ge&I(X;U_i^{m-1})\\
R_0&\ge&I(X,Y_1^{m-1};U_1^{m-1})
\end{array}},\label{result2}\end{equation}where $D_m$ is the set of distributions
\begin{equation}\setbr{\begin{array}{r c l}
P_{X,Y_1^{m-1},U_1^{m-1}}&:& \forall~1\le i\le m-1,~1\le j\le m-2,\\
&&(X,Y_1^{m-1})\sim Q_XQ_{Y_1^{m-1}|X},\\
&&X-U_1^{m-1}-Y_1,\\
&&(X,Y_1^j,U_1^j)-U_{j+1}^{m-1}-Y_{j+1},\\
&&H(U_{i}^{m-1}|U_i)=0,
\end{array}}\end{equation}
with the cardinality bounds
\begin{equation}
\abs{\U_i}\le\abs{\X}\br{\prod_{k=1}^{m-1}\abs{\Y_k}}\br{\prod_{k=i+1}^{m-1}\abs{\U_k}}+m+i-2,
\label{gencard}\end{equation}
for $1\le i\le m-1$. The proof is similar to the proof of Theorem 1, which is presented in sections IV and V.  The appendix contains a proof outline for the general case.

\section{Variations on Cascade Channel Synthesis}

General solutions for modifications of the problem considered above remain elusive.  However, we attempt to derive some qualitative insights by considering two simple variations in this section.

\subsection{No Secrecy on Link 1}

Recall that the channel synthesis problem for the cascade setting with no secrecy constraints \cite{Bloch3} is unsolved.  We now demonstrate how the rate region expands when the secrecy constraint is relaxed.
Consider the network of Fig. \ref{sccs}, with the first node's message unseen by the eavesdropper.  In this case, we modify our definition of achievability \eqref{synthdef} to a sequence of codes that satisfy
\begin{equation}
\lim_{n\to\infty}\normtv{P_{X^nY^nZ^nM_2}-P_{M_2}\prod Q_{XYZ}}=0.\label{locsec}
\end{equation}
Due to the relaxed secrecy criterion, some of the burden carried by the common randomness can be distributed to communication on the first link.  In fact, it may be beneficial to discard the superposition structure and have the second node perform local actions to generate its message.

Consider the rate regions
\begin{equation}\s_{D}^{(in)}\defeq \setbr{\begin{array}{r c l}
\hspace{-.1cm}R_0^2\in\R^3&:&\ex P_{X,Y,Z,U,V}\in D \mbox{ s.t.}\\
R_1&\ge&I(X;U,V)\\
R_2&\ge&I(X;V)\\
R_0&\ge&I(X,Y,Z;V)\\
R_1+R_0&\ge&I(X,Y,Z;U,V)+I(X;V)
\end{array}},\label{locsecin}\end{equation}
and
\begin{equation}\s_{D}^{(out)}\defeq \setbr{\begin{array}{r c l}
\hspace{-.1cm}R_0^2\in\R^3&:&\ex P_{X,Y,Z,U,V}\in D \mbox{ s.t.}\\
R_1&\ge&I(X;U,V)\\
R_2&\ge&I(X;V)\\
R_0&\ge&I(X,Y,Z;V)\\
R_1+R_0&\ge&I(X,Y,Z;U,V)
\end{array}},\label{locsecout}\end{equation}
where $D$ is defined in \eqref{D}.

\begin{theorem}
Let $\C_{loc1}$ denote the closure of the set of achievable rates under the local secrecy criterion \eqref{locsec}.  We have
\begin{equation}
\s_{D}^{(in)}\subseteq \C_{loc1}\subseteq \s_{D}^{(out)}.
\end{equation}
\end{theorem}

As shown in the appendix, the result is tight for any distribution which can be efficiently synthesized under the constraint $I(Y;U|V,X)=0$.  The task assignment example, where the number of tasks equals the number of nodes in the cascade ($m=3$), is an example where the above bounds are tight.  The rate region is given by all rate tuples in

\begin{equation}\setbr{\begin{array}{r c l}
R_1&\ge&\log3\\
R_2&\ge&\log3-1\\
R_0&\ge&\log3
\end{array}},\end{equation}
which yields a 1 bit discount in the rate of common randomness when compared to
\begin{equation}\setbr{\begin{array}{r c l}
R_1&\ge&\log3\\
R_2&\ge&\log3-1\\
R_0&\ge&\log3+1
\end{array}},\end{equation}
the rate region given by \eqref{task}.  Note that the projection of the region onto the communication rates $(R_1,R_2)$ is invariant (cf.\ \cite[Corollary 1]{Bloch3} and \cite[Theorem 2]{coord}).
Proofs are presented in the appendix.

\subsection{No Secrecy on Link 2}

For completeness, we also consider the case where only the transmission on the first link is to be secured.  Note that the optimal scheme for the fully secure cascade synthesis problem reduces to concealing the transmission on the first link.  However, we show here that the solution for this relaxed problem remains elusive.  The gap is the functional relation between auxiliary random variables, that was established in the fully secure problem.

Consider the network of Fig. \ref{sccs}, with the second node's message unseen by the eavesdropper.  We modify our definition of achievability \eqref{synthdef} to a sequence of codes that satisfy
\begin{equation}
\lim_{n\to\infty}\normtv{P_{X^nY^nZ^nM_1}-P_{M_1}\prod Q_{XYZ}}=0.\label{locsec1}
\end{equation}
Again, it may be beneficial to discard the superposition structure and have the second node perform local actions to generate its message.  However, we present the achievable region used in the fully secure scheme to illustrate the gap that arises in the analysis.

Consider the rate region
\begin{equation}\s_{D}^{(loc)}\defeq  \setbr{\begin{array}{r c l}
\hspace{-.1cm}(R_0,R_1,R_2)\in\R^3&:&\ex P_{X,Y,Z,U,V}\in D \mbox{ s.t.}\\
R_1&\ge&I(X;U)\\
R_2&\ge&I(X;V)\\
R_0&\ge&I(X,Y,Z;U)
\end{array}},\label{locsec1reg}\end{equation}
where $D$ is defined in \eqref{D}.  Note that $\s_{D'}^{(loc)}$ is the region $\s_{D'}$ in Theorem 1.

\begin{theorem}
Let $\C_{loc2}$ denote the closure of the set of achievable rates under the local secrecy criterion \eqref{locsec1}.  We have
\begin{equation}
\s_{D'}^{(loc)}\subseteq \C_{loc2}\subseteq \s_{D}^{(loc)}.
\end{equation}
\end{theorem}

Recall that $D'=D\cap\setbr{P_{X,Y,Z,U,V}:H(V|U)=0}$, so the functional relation established in Theorem 1 is really particular to the fully secure cascade synthesis problem.  Proofs are presented in the appendix.

\subsection{Cascade with Relay}

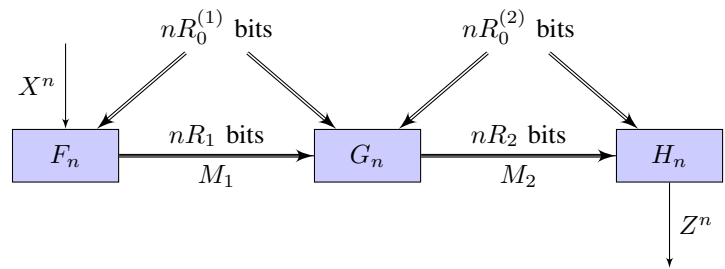
\begin{figure}[ht]
\begin{center}
\resizebox{3.7in}{1.4in}{
\begin{tikzpicture}[scale=1,auto, node distance=1cm,>=latex']
    \node [input, name=input] {};
    \node [block, below of=input, node distance=1.5cm] (node1) {$F_n$};
    \node [block, right of=node1,
            node distance=4cm] (node2) {$G_n$};
    \node [block, right of=node2,
            node distance=4cm] (node3) {$H_n$};
    \draw [->,double] (node2) -- node[name=r2] {$nR_2$ bits} (node3);
    \draw [] (node3) -- node[name=j2] {$M_2$} (node2);
    \node [draw=none, above of=j2, node distance = 2cm] (commrand1) {$nR_0^{(2)}$ bits};
    \node [output, below of=node2, node distance=1.5cm] (output1) {};
    \node [output, below of=node3, node distance=1.5cm] (output2) {};

    \draw [draw,<-] (node1) -- node {$X^n$} (input);
    \draw [->,double] (node1) -- node {$nR_1$ bits} (node2);
    \draw [] (node2) -- node [name=mid1] {$M_1$} (node1);
    \node [draw=none, above of=mid1, node distance = 2cm] (commrand) {$nR_0^{(1)}$ bits};
    \draw [->] (node3) -- node [name=zn] {$Z^n$}(output2);
    \draw [->,double] (commrand) -- (node1);
    \draw [->,double] (commrand) -- (node2);
    \draw [->,double] (commrand1) -- (node2);
    \draw [->,double] (commrand1) -- (node3);
\end{tikzpicture}}
\caption{The i.i.d.\ sequence $X^n$ is given by nature. Messages $M_1$ and $M_2$ are sent along the cascade at rates $R_1$ and $R_2$. Common randomness $K_1$ and $K_2$ are shared by adjacent nodes at rates $R_0^{(1)}$ and $R_0^{(2)}$. We want $(X^n,Z^n)$ i.i.d.\ correlated and independent of the messages $(M_1,M_2)$.}\label{sccsrelay}
\end{center}
\end{figure}

Provision of a centralized source of common randomness seems like an optimistic assumption for many practical scenarios.  A more realistic assumption is common randomness shared between adjacent nodes of a communication network.  Consider the cascade network in Fig.\ \ref{sccsrelay}, endowed with this structure.

For simplicity, we set $Y=\es$ so that the intermediate node of the cascade plays the passive role of a relay.  Furthermore, we shall assume that we have an abundant supply of common randomness on both links.  This renders any secrecy requirement superfluous.  We wish to produce a sequence $Z^n$ at the final node of the cascade such that
\begin{equation}
\lim_{n\to\infty}\normtv{P_{X^nZ^n}-\prod Q_{XZ}}=0,\label{relay}
\end{equation}
which is a simplification of \eqref{synthdef}.  The question we seek to address is:  is there a communication-cost of replacing a centralized source of randomness with localized common randomness?

Conjecture 1 of \cite{coord} raises the question of whether optimal strategies for empirical coordination can be converted into optimal strategies for strong coordination, simply by augmenting the scheme with sufficient common randomness.  The following result provides a negative answer if the common randomness is not shared among all nodes.  Note that the conjecture remains an intriguing open question for networks with a single, centralized source of common randomness.

\begin{theorem}
Let $\C_{relay}$ denote the closure of the set of achievable communication rates under the coordination criterion \eqref{relay}.  We have
\begin{equation}
\C_{relay}=\s_{relay},
\end{equation}
where
\begin{equation}\s_{relay}\defeq \setbr{\begin{array}{r c l}
\hspace{-.1cm}(R_1,R_2)\in\R^2&:&\ex P_{X,Z,U}\in D_{relay} \mbox{ s.t.}\\
R_1&\ge&I(X;U)\\
R_2&\ge&I(Z;U)
\end{array}},\label{relayresult}\end{equation}and
\begin{equation}D_{relay}\defeq\setbr{\begin{array}{r c l}
P_{X,Z,U}&:&(X,Z)\sim Q_XQ_{Z|X},\\
&&X-U-Z,\\
&&\abs{\U}\le\abs{\X}+\abs{\Z}+2
\end{array}}.\label{relayD}\end{equation}
\end{theorem}

The proofs for this section are provided in the appendix.

Empirical coordination on the cascade network, i.e.\ synthesis of $Z^n$ that is jointly typical with $X^n$, was studied in \cite[Theorem 5]{coord}.  The optimal empirical coordination scheme achieves $R_1^{(emp)}=R_2^{(emp)}=I(X;Z)$.  We illustrate the difference with the above region by example.

Consider the scatter channel \cite[II-H]{DCS}, where $(X,Z)$ is drawn uniformly from all pairs $(X,Z)\in[m]^2$ with $X\ne Z$ and $m\ge 2$.  By arguments similar to the evaluation of \eqref{result} for the task assignment example in section II-E, \eqref{relayresult} simplifies to
\begin{equation}
\C_{relay}=\mbox{Conv}\br{\setbr{\begin{array}{r c l}
\hspace{-.1cm}(R_1,R_2)\in\R^2&:&\ex a\in[m-1] \mbox{ s.t.}\\
R_1&\ge&\log\br{\frac{m}{a}}\\
R_2&\ge&\log\br{\frac{m}{m-a}}
\end{array}}}.\label{relaytask}
\end{equation}

The optimal empirical coordination scheme achieves
\begin{equation}
R_1^{(emp)}=R_2^{(emp)}=I(X;Z)=\log\frac{m}{m-1}.
\end{equation}
On the other hand, setting $a=m-1$ to achieve $R_1=\log\br{\frac{m}{m-1}}$ implies that $R_2=\log m$.  The difference $R_2-R_2^{(emp)}=\log(m-1)$ is unbounded, as a function of $m$.  On the other hand, the optimal sum-rate in \eqref{relayresult} is about 2 bits (exact for even $m$).


\section{Achievability Proof of Theorem 1}

\subsection{Total Variation Properties}

Before delving into the achievability proof, we list some useful properties of the total variation distance:
\begin{itemize}
\item Total variation between marginals is upper-bounded by the total variation between joint distributions \cite[Lemma V.1]{DCS}, i.e.
\begin{equation}
\normtv{P_X-Q_X}\le \normtv{P_{X,Y}-Q_{X,Y}}.
\label{v.1}
\end{equation}

\item Total variation between joint distributions reduces to total variation between marginals if conditional distributions given the marginal variables coincide \cite[Lemma V.2]{DCS}, i.e.\
\begin{equation}
\normtv{P_XP_{Y|X}-Q_XP_{Y|X}}= \normtv{P_{X}-Q_{X}}.
\label{v.2}
\end{equation}

\item The expected value of a bounded function is continuous w.r.t.\ total variation i.e.\
\begin{equation}
\abs{\E_Pf(X)-\E_Qf(X)}\le 2f_{max} \normtv{P-Q},
\label{tvcont}
\end{equation}
where $f_{max}\defeq \max_{x\in\X}\abs{f(x)}$.

\end{itemize}

\subsection{Soft-Covering}

Our random number generation scheme is hinged on a distributional approximation result, which we refer to as Wyner's soft-covering lemma \cite{DCS}.  It is implied by results on resolvability \cite{resolvability} and goes back to Wyner's work on common information \cite{Wyner}.  For literature about related extensions, readers may refer to \cite{DCS}.

Here is the most basic setting that the lemma addresses. Given a memoryless channel $Q_{X|U}$, we want to synthesize $X^n\sim\prod Q_X$ at the output. However, we would like to do it using an input selected randomly from a small codebook of $U^n\sim\prod Q_U$ codewords. How large does the codebook need to be?  The lemma provides a sufficient condition (that is also necessary \cite{resolvability}) in order to avoid a biased $X^n$ sequence.

\begin{lemma}[Lemma IV.1 in  \cite{DCS}]
\label{soft}
For any discrete distribution $ Q_{UX}$, let $\B^{(n)}=\setbr{U^n(m)}_{m=1}^{2^{nR}}$ be a codebook of sequences each drawn independently from $\prod Q_U$. For a fixed codebook, define\vspace{-0.2cm}
\begin{equation}
P_{X^n}=\frac{1}{2^{nR}}\sum_{m=1}^{2^{nR}}\prod_{k=1}^n Q_{X|U}(x_k|U_k(m)),\vspace{-0.2cm}
\end{equation}
the distribution induced by passing a random $U^n$ codeword through the memoryless channel $Q_{X|U}$.
Then, $R>I(X;U)$ implies that \vspace{-0.1cm}
\begin{equation}
\E\normtv{P_{X^n}-\prod_{k=1}^n Q_X(x_k)}<\e_n,\vspace{-0.1cm}
\end{equation}
with $\lim_{n\to\infty}\e_n=0$, where the expectation is w.r.t.\ ${\cal B}^{(n)}$.
\end{lemma}

Since our synthesis scheme is based on a superposition code, we will need a generalized version of Lemma \ref{soft}, provided by the ``generalization of Lemma 6.1 of [5]" in \cite{Gohari1}.  A closely related result is the superposition soft-covering lemma \cite[Corollary VII.8]{DCS}.

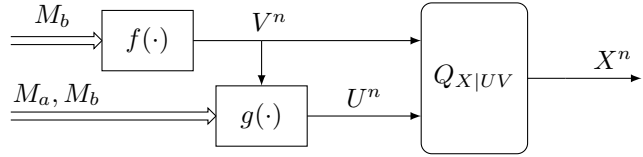
\begin{figure}[ht]
\begin{center}
\begin{tikzpicture}[node distance=2cm]
 \node (src1)   [coordinate] {};
 \node (src2)   [coordinate,below=1cm of src1] {};
 \node (chv)    [coordinate,below=5mm of src1] {};
 \node (label1) [coordinate,right=2cm of src1] {};
 \node (enc1)   [node,minimum width=12mm,right=12mm of src1] {$f(\cdot)$};
 \node (enc2)   [node,minimum width=12mm,right=27mm of src2] {$g(\cdot)$};
 \node (split1) [coordinate] at (src1 -| enc2.center) {};
 \node (chh)    [coordinate,right=22mm of enc2] {\distnarg[\Phi]{V|U}};
 \node (ch)     [node,minimum width=14mm,minimum height=2cm,rounded corners] at (chv -| chh) {\distnarg[Q]{X|UV}};
 \node (dashline) [coordinate,right=5mm of ch] {};
 \node (dash1)  [coordinate] at (src1 -| dashline) {};
 \node (dash2)  [coordinate] at (src2 -| dashline) {};
 \node (outline)  [coordinate,right=1cm of dashline] {};
 \node (out1)   [coordinate] at (src1 -| outline) {};
 \node (out2)   [coordinate] at (src2 -| outline) {};

 \rightratearrow{src1}{enc1.west}{1.5pt}{3pt}{$M_b$}{3pt};
 \draw[draw=none] (src2) to node [midway,above] {$M_a,M_b$} (src2 -| enc1.west);
 \rightratearrow{src2}{enc2.west}{1.5pt}{3pt}{}{3pt};
 \draw (enc1) to node [midway,above] {~~~~~~~~~$V^n$} (split1);
 \draw[arw] (split1) to (src1 -| ch.west);
 \draw[arw] (split1) to (enc2);
 \draw[arw] (enc2) to node [midway,above] {$U^n$} (src2 -| ch.west);
 \draw (ch) to (ch.center -| dashline);
 \draw[arw] (chv -| dashline) to node [midway,above] {~~$X^n$} (chv -| out2);
\end{tikzpicture}
\caption{{\em Generalized Soft-Covering Lemma:}  Codewords $(U^n,V^n)$ drawn randomly from a superposition codebook are passed through the memoryless channel $Q_{X|UV}$.}
\label{gensoftfig}
\end{center}
\end{figure}

\begin{lemma}[Generalized Soft-Covering \cite{Gohari1}]
\label{gensoft}

For any discrete distribution $ Q_{XUV}$, let ${\cal B}_{V^n}^{(n)}=\setbr{v^n(m_b)}_{m_b=1}^{2^{nR_b}}$ be a codebook of sequences each drawn independently from $\prod \distnarg[Q]{V}$.  For each $v^n(m_b)\in{\cal B}_{V^n}^{(n)}$, let ${\cal B}_{U^n|V^n}^{(n)}=\setbr{u^n(m_a,m_b)}_{m_a=1}^{2^{nR_a}}$ be a codebook of sequences each drawn independently from $\prod \distnarg[Q]{U|V}$.  For fixed codebooks, define
\begin{equation}
\distnarg{X^n}=\frac{1}{2^{n(R_a+R_b)}}\sum_{m_a=1}^{2^{nR_a}}\sum_{m_b=1}^{2^{nR_b}}\prod_{k=1}^n Q_{X|VU}(x_k|V_k(m_b),U_k(m_a,m_b)),
\end{equation}
the conditional distribution induced by passing random $(U^n,V^n)$ codewords through the memoryless channel $Q_{X|UV}$, as shown in Fig.~\ref{gensoftfig}.  Then,
\begin{align}
R_b&> I(X;V),\\
R_a+R_b&> I(X;U,V),
\end{align}
imply that
\begin{equation}
\E\normtv{P_{X^n}-\prod_{k=1}^n Q_X(x_k)}<\delta_n,\label{geneq}
\end{equation}
with $\lim_{n\to\infty}\delta_n=0$, where the expectation is w.r.t.\ ${\cal B}_{V^n}^{(n)}$ and ${\cal B}_{U^n|V^n}^{(n)}$.
\end{lemma}

Note that the lemma places a stricter constraint on the base layer ${\cal B}_{V^n}^{(n)}$ of the superposition code.  For example, if the $V^n$ codebook is too small, then we will have a biased sample of channel instances $\prod Q_{X|U,V=v_k}$, which will hinder synthesis of $X^n$ even if ${\cal B}_{U^n|V^n}^{(n)}$ is extremely large.  The lemma provides a sufficient condition to avoid this bias.


\subsection{Construction of Idealized Distribution}

Assume that $(R_0,R_1,R_2)$ is in the interior of $\s_D$. Then there exists a distribution $ Q_{XYZUV}\in D$ such that the rates in \eqref{result} are strictly satisfied. We now describe an idealized distribution $\Upsilon$, from which we shall derive our encoders.  This idealized distribution acts as a simple proxy for the actual coding scheme.

For $n\ge1$, let $(K,M_a,M_b)$ be uniformly distributed on $[2^{nR_0}]\times[2^{n(R_1-R_2)}]\times[2^{nR_2}]$.  We shall set $M_1:=(M_a,M_b)$ and $M_2:=M_b$.  Consider a codebook $\B_{V^n}$ of $2^{n(R_0+R_2)}$ $V^n$ sequences (randomly drawn according to the i.i.d.\ distribution $\prod Q_V$), where the codewords are indexed as $v^n(M_b,K)$.  For every $V^n$ codeword, we have a codebook $\B_{U^n|V^n}$ of $2^{n(R_1-R_2)}$ $U^n$ sequences (randomly drawn by passing the $V^n$ codeword through the memoryless channel $\prod Q_{U|V}$), indexed as $u^n(M_a,M_b,K)$.
\begin{figure}[ht]
\begin{center}
\begin{tikzpicture}[node distance=2cm]
 \node (src1)   [coordinate] {};
 \node (src2)   [coordinate,below=1cm of src1] {};
 \node (chv)    [coordinate,below=5mm of src1] {};
 \node (label1) [coordinate,right=2cm of src1] {};
 \node (enc1)   [node,minimum width=12mm,right=12mm of src1] {$f(\cdot)$};
 \node (enc2)   [node,minimum width=12mm,right=27mm of src2] {$g(\cdot)$};
 \node (split1) [coordinate] at (src1 -| enc2.center) {};
 \node (chh)    [coordinate,right=22mm of enc2] {\distnarg[\Phi]{V|U}};
 \node (ch)     [node,minimum width=14mm,minimum height=2cm,rounded corners] at (chv -| chh) {\distnarg[Q]{XYZ|UV}};
 \node (dashline) [coordinate,right=5mm of ch] {};
 \node (dash1)  [coordinate] at (src1 -| dashline) {};
 \node (dash2)  [coordinate] at (src2 -| dashline) {};
 \node (outline)  [coordinate,right=1cm of dashline] {};
 \node (out1)   [coordinate] at (src1 -| outline) {};
 \node (out2)   [coordinate] at (src2 -| outline) {};

 \rightratearrow{src1}{enc1.west}{1.5pt}{3pt}{$M_b,K$}{3pt};
 \draw[draw=none] (src2) to node [midway,above] {$M_a,M_b,K$} (src2 -| enc1.west);
 \rightratearrow{src2}{enc2.west}{1.5pt}{3pt}{}{3pt};
 \draw (enc1) to node [midway,above] {~~~~~~~~~$V^n$} (split1);
 \draw[arw] (split1) to (src1 -| ch.west);
 \draw[arw] (split1) to (enc2);
 \draw[arw] (enc2) to node [midway,above] {$U^n$} (src2 -| ch.west);
 \draw (ch) to (ch.center -| dashline);
 \draw[arw] (chv -| dashline) to node [midway,above] {~~$X^n,Y^n,Z^n$} (chv -| out2);
\end{tikzpicture}
\caption{{\em Auxiliary Idealized Distribution $\Upsilon$:}  Codewords $(U^n,V^n)$ drawn randomly from a superposition codebook are passed through the memoryless channel $Q_{XYZ|UV}$.}
\label{ideal}
\end{center}
\end{figure}
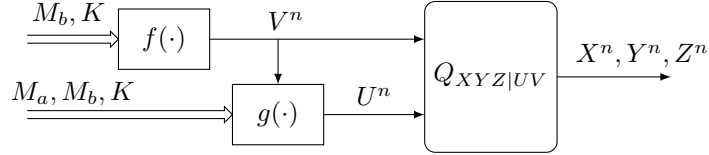

Note that the Markov chains $X-(U,V)-Y$ and $(X,Y,U)-V-Z$ imply that $Q_{XYZ|UV}=Q_{X|UV}Q_{Y|UV}Q_{Z|V}$, so the memoryless channel decouples, yielding Markov chains $X^n-U^nV^n-Y^n$ and $X^nY^nU^n-V^n-Z^n$ under the idealized distribution $\Upsilon$.  Precisely, we have 
\begin{align}
&~\quad\Up_{X^n,Y^n,Z^n,M_a,M_b,K}\defeq\frac{1}{2^{n(R_0+R_1)}}\br{\prod_{t=1}^n Q_{XYZ|UV}(x_t,y_t,z_t|u_t(\underbrace{m_a,m_b}_{m_1},k),v_t(\underbrace{m_b}_{m_2},k))}.
\end{align}

Note that both Markov chains $X^n-M_1K-M_2Y^n$ and $X^nY^nM_1-M_2K-Z^n$ are satisfied, consistent with the physical constraints of cascade communication. Finally, we set the operational distribution (our synthesis scheme) to be defined by
\begin{align}
P_{X^n,Y^n,Z^n,M_a,M_b,K}&\defeq\frac{1}{2^{nR_0}}\br{\prod Q_X}\Up_{Y^n,Z^n,M_a,M_b|X^n,K}.\label{scheme}
\end{align}
The first node picks messages according to $\Up_{M_a,M_b|X^n,K}$, the second node passes $u^n(M_a,M_b,K)$ through the memoryless channel $\prod Q_{Y|UV}$ and the final node passes $v^n(M_b,K)$ through the memoryless channel $\prod Q_{Z|V}$.

\subsection{Total Variation Analysis}

We now proceed to show that there exist codebooks such that the above construction \eqref{scheme} meets the secure synthesis criterion \eqref{synthdef}.  First, note that by the triangle inequality and \eqref{v.1}, we have (expectation is over the $(U^n,V^n)$ codebook)
\begin{align}
\E\normtv{P_{X^nY^nZ^nM_1M_2}-P_{M_1M_2}\prod Q_{XYZ}}
&\le \E\normtv{P_{X^nY^nZ^nM_1M_2}-\Up_{X^nY^nZ^nM_1M_2}}+\cdots\nonumber\\
&\quad~~\E\normtv{\Up_{X^nY^nZ^nM_1M_2}-\Up_{M_1M_2}\prod Q_{XYZ}}+\cdots\nonumber\\
&\quad~~\E\normtv{\Up_{M_1M_2}\prod Q_{XYZ}-P_{M_1M_2}\prod Q_{XYZ}}\\
&= \E\normtv{P_{X^nY^nZ^nM_1M_2}-\Up_{X^nY^nZ^nM_1M_2}}+\cdots\nonumber\\
&\quad~~\E\normtv{\Up_{X^nY^nZ^nM_1M_2}-\Up_{M_1M_2}\prod Q_{XYZ}}+\cdots\nonumber\\
&\quad~~\E\normtv{\Up_{M_1M_2}-P_{M_1M_2}}\\
&\le 2\E\normtv{P_{X^nY^nZ^nM_1M_2}-\Up_{X^nY^nZ^nM_1M_2}}+\cdots\nonumber\\
&\quad~~\E\normtv{\Up_{X^nY^nZ^nM_1M_2}-\Up_{M_1M_2}\prod Q_{XYZ}}.\label{inter}
\end{align}

Thus, we have reduced the problem to showing that
\begin{itemize}
\item the idealized distribution $\Up$ satisfies \eqref{synthdef}, and
\item the operational distribution $P$ is well-approximated by $\Up$.
\end{itemize}

We address the former first.  Note that with fixed communication $(m_a,m_b)$, the scheme may choose from $2^{nR_0}$ $W^n$ codewords, where $W\defeq (U,V)$. By Lemma \ref{soft}, $R_0>I(X,Y,Z;W)=I(X,Y,Z;U,V)$ implies that for any $(m_a,m_b)\in[2^{n(R_1-R_2)}]\times[2^{nR_2}]$, we have
\begin{equation}
\E\normtv{\Up_{X^nY^nZ^n|M_1=(m_a,m_b),M_2=m_b}-\prod Q_{XYZ}}<\e_n,
\end{equation}
with $\lim_{n\to\infty}\e_n=0$.  This implies that
\begin{align}
\E\normtv{\Up_{X^nY^nZ^nM_1M_2}-\Up_{M_1M_2}\prod Q_{XYZ}}
&=\E\frac{1}{2}\sum_{x^ny^nz^nm_1m_2}\abs{\Up_{X^nY^nZ^nM_1M_2}-\frac{1}{2^{nR_1}}\prod Q_{XYZ}}\\
&=\frac{1}{2^{nR_1}}\E\frac{1}{2}\sum_{x^ny^nz^nm_1m_2}\abs{\Up_{X^nY^nZ^n|M_1M_2}-\prod Q_{XYZ}}\\
&=\frac{1}{2^{nR_1}}\sum_{m_1m_2}\E\normtv{\Up_{X^nY^nZ^n|M_1M_2}-\prod Q_{XYZ}}\\
&<\frac{1}{2^{nR_1}}\sum_{m_1m_2}\e_n\\
&=\e_n,\label{ach1}
\end{align}
so indeed, $\Up$ satisfies \eqref{synthdef}.

Finally, we have to ensure that $P$ is well-approximated by $\Up$. Consider the first term in \eqref{inter} - by \eqref{v.1} and \eqref{v.2}, we have
\begin{align}
\E\normtv{P_{X^nY^nZ^nM_1M_2}-\Up_{X^nY^nZ^nM_1M_2}}
&\le \E\normtv{P_{X^nY^nZ^nM_1M_2K}-\Up_{X^nY^nZ^nM_1M_2K}}\\
&= \E\normtv{P_{X^nK}-\Up_{X^nK}},\label{inter1}
\end{align} 
since $P_{Y^nZ^nM_1M_2|X^nK}=\Up_{Y^nZ^nM_1M_2|X^nK}$ by definition \eqref{scheme}.  Recall that $P_{X^nK}=2^{-nR_0}\prod Q_X$, by the problem definition. Thus, it remains to argue that the idealized distribution $\Up$ in Fig. \ref{ideal} generates marginally i.i.d.\ $X^n$ that is independent of $K$.  

Note that with fixed common randomness $k$, the scheme may choose from $2^{nR_1}$ $W^n$ codewords, where $W\defeq (U,V)$. Since the codebook has a superposition structure, by Lemma \ref{gensoft}, we have that ($R_a:=R_1-R_2$ and $R_b:=R_2$)
\begin{align}
R_2&>I(X;V),\\
R_1&>I(X;U,V),
\end{align}
imply that for any $k\in[2^{nR_0}]$, we have
\begin{equation}
\E\normtv{\Up_{X^n|K=k}-\prod Q_{X}}<\delta_n,
\end{equation}
with $\lim_{n\to\infty}\delta_n=0$.  This implies that

\begin{align}
\E\normtv{P_{X^nK}-\Up_{X^nK}}&=\E\sum_{x^n,k}\frac{1}{2}\abs{\frac{1}{2^{nR_0}}\prod Q_X -\frac{1}{2^{nR_0}}\Up_{X^n|K=k}}\\
&=\frac{1}{2^{nR_0}}\sum_{k}\E\normtv{\prod Q_X -\Up_{X^n|K=k}}\\
&<\frac{1}{2^{nR_0}}\sum_{k}\delta_n\\
&=\delta_n.\label{inter2}
\end{align}
Combining \eqref{inter}, \eqref{ach1}, \eqref{inter1} and \eqref{inter2}, for $n$ sufficiently large, we have for any $\e>0$ that
\begin{equation}
\E\normtv{P_{X^nY^nZ^nM_1M_2}-P_{M_1M_2}\prod Q_{XYZ}}<\e.
\end{equation}
Thus, for all $n$ sufficiently large, there must exist a choice of $\B_{V^n}$ and $\B_{U^n|V^n}$ codebooks that deterministically achieve the above bound, with the rate constraints specified in \eqref{result}.


\subsection{Comment on Achievability}
Our scheme requires local randomization at all nodes, including stochastic encoders - also known as the \emph{likelihood} encoder \cite{eva}.  Please refer to  \cite{DCS, Vell1} for a quantitative treatment of local randomness in channel synthesis. Also, observe that while it is intuitive to think of the common randomness as a one-time pad on the messages, we do not need to use such a construction in our proof. On the other hand, it may be desirable to have a more direct achievability scheme for channel synthesis with explicit constructions. Some attempts have been made in this direction \cite{Gohari5,Bloch1}.

\section{Converse Proof of Theorem 1}

Let $(R_0,R_1,R_2)$ be achievable. Then for $\e\in(0,1/4)$ there exists a $(2^{nR_0},2^{nR_1},2^{nR_2},n)$ secure channel synthesis code with an induced joint distribution $P_{X^n,Y^n,Z^n,K,M_1,M_2}$ such that
\begin{equation}
\normtv{P_{X^nY^nZ^nM_1M_2}-P_{M_1M_2}\prod Q_{XYZ}}<\e,
\end{equation}
for $n$ sufficiently large.
First, we use the triangle inequality and  \eqref{v.1} to note that
\begin{align}
\normtv{P_{X^nY^nZ^nM_1M_2}-P_{M_1M_2}P_{X^nY^nZ^n}}
&\le\normtv{P_{X^nY^nZ^nM_1M_2}-P_{M_1M_2}\prod Q_{XYZ}}+\ldots\nonumber\\
&\qquad\normtv{P_{M_1M_2}P_{X^nY^nZ^n}-P_{M_1M_2}\prod Q_{XYZ}}\\
&=\normtv{P_{X^nY^nZ^nM_1M_2}-P_{M_1M_2}\prod Q_{XYZ}}+\ldots\nonumber\\
&\qquad\normtv{P_{X^nY^nZ^n}-\prod Q_{XYZ}}\\
&\le2\normtv{P_{X^nY^nZ^nM_1M_2}-P_{M_1M_2}\prod Q_{XYZ}}<2\e.\vspace{-.1cm}\label{indep}\end{align}

Theorem 17.3.3 of  \cite{Cover} coupled with \eqref{indep} implies that
\begin{align}
I(X^nY^nZ^n;M_1M_2)
&=H(X^nY^nZ^n)+H(M_1M_2)-H(X^nY^nZ^nM_1M_2)\\
&< n\e (\log\br{\abs{\X}\abs{\Y}\abs{\Z}}+R_1+R_2)-\e\log\e\\
&:=n g_1(\e),\label{g1}
\end{align}
where $g_1(\e)$ is defined by the above equality.  Note that $\lim_{\e\downarrow0}g_1(\e)=0$.

\subsection{Entropy Bounds}

We will need the following bounds on entropy in terms of total variation \cite[Lemma VI.3]{DCS}. If the joint distribution of $(X^n,Y^n,Z^n)$ is close in total variation to an i.i.d.\ distribution as assumed, then we have
\begin{align}
\sum_{t=1}^nI_P(X_t,Y_t,Z_t;X^{t-1},Y^{t-1},Z^{t-1})&\le ng(\e),\label{tv1}\\
I_P(X_T,Y_T,Z_T;T)&\le ng(\e),\label{tv2}
\end{align}
where
\begin{equation}
g(\e)\defeq 4\e\log\br{\frac{\abs{\X}\abs{\Y}\abs{\Z}}{\e}}.
\end{equation}
Note that $\lim_{\e\downarrow0}g(\e)=0$.

We shall use the random variable $T$ uniformly distributed on $[n]$, as a random time index.  We also need the result \cite[Lemma VI.2]{DCS} that for distributions $P_{X^n}$ and $Q_{X^n}$, we have
\begin{equation}
\normtv{P_{X_T}-Q_{X_T}}\le \normtv{P_{X^n}-Q_{X^n}}\label{vi.2}
\end{equation}

\subsection{Approximate Rate Region}

We use standard information-theoretic inequalities, the physical constraint $X^n-(M_1,K)-M_2$ and the fact that $X^n$ is i.i.d.\ and independent of $K$ to bound the communication rates:\vspace{-.2cm}
\begin{multicols}{2}
\begin{align}
nR_1&\ge H(M_1)\\
&\ge H(M_1|K)\\
&\ge I(X^n;M_1|K)\\
&= I(X^n;M_1,M_2|K)\\
&=I(X^n;M_1,M_2,K)\\
&=\sum_{i=1}^nI(X_i;M_1,M_2,K|X^{i-1})\\
&=\sum_{i=1}^nI(X_i;M_1,M_2,K,X^{i-1})\\
&\ge\sum_{i=1}^nI(X_i;M_1,M_2,K)\\
&=nI(X_T;M_1,M_2,K|T)\\
&=nI(X_T;M_1,M_2,K,T),
\end{align}

\columnbreak

\begin{align}
nR_2&\ge H(M_2)\\
&\ge H(M_2|K)\\
&\ge I(X^n;M_2|K)\\
&=I(X^n;M_2,K)\\
&=\sum_{i=1}^n I(X_i;M_2,K|X^{i-1})\\
&\ge \sum_{i=1}^n I(X_i;M_2,K)\\
&= n I(X_T;M_2,K|T)\\
&= n I(X_T;M_2,K,T).
\vspace{-.2cm}
\end{align}
\end{multicols}
Finally, we bound $R_0$:\vspace{-.2cm}
\begin{align}
nR_0
&\ge H(K)\\
&\ge H(K|M_1,M_2)\\
&\ge I(X^n,Y^n,Z^n;K|M_1,M_2)\\
&\ge I(X^n,Y^n,Z^n;M_1,M_2,K)-ng_1(\e)\label{g1use}\\
&\ge\sum_{t=1}^nI(X_t,Y_t,Z_t;M_1,M_2,K)-ng_1(\e)-ng(\e)\\
&\ge nI(X_T,Y_T,Z_T;M_1,M_2,K|T)-n(g_1(\e)+g(\e))\\
&\ge nI(X_T,Y_T,Z_T;M_1,M_2,K,T)-n(g_1(\e)+2g(\e)).\label{corr}
\end{align}
The inequality \eqref{g1use} follows from \eqref{g1}, while the other steps follow from \eqref{tv1} and \eqref{tv2}. Making associations $(X_T,Y_T,Z_T)=(X,Y,Z)$, $U=M_1$ and $V=(M_2,K,T)$, we see that the rates and Markov chains in \eqref{result} and \eqref{D} are satisfied up to the correction in \eqref{corr}. 

Using the Carath\'{e}odory theorem for connected sets \cite{egg}, we can show that $\abs{\V}\le\abs{\X}\abs{\Y}\abs{\Z}+3$ suffices to ensure the existence of a distribution $P'_{XYZUV}$ that preserves $H(X|U,V)$, $H(X|V)$, $H(X,Y,Z|U,V)$, $I(X,Y,U;Z|V)$ and the marginal on $(X,Y,Z)$.  Note that preserving $I(X,Y,U;Z|V)$ retains the Markov chain $(X,Y,U)-V-Z$. The resulting distribution is formed by an average over distributions of $(X,Y,Z,U)$ that satisfy $I(X,Y,U;Z)=0$ and $X-U-Y$, thus preserving both Markov chains in \eqref{D}.

Next, we apply the Carath\'{e}odory theorem again to argue that $\abs{\U}\le\abs{\X}\abs{\Y}\abs{\Z}\abs{\V}+3$ suffices to ensure the existence of a distribution $\Gamma_{XYZUV}$ that preserves $H(X|U,V)$, $H(X,Y,Z|U,V)$, $H(Z|X,Y,U,V)$, $I(X;Y|U,V)$ and the marginal on $(X,Y,Z,V)$.  Preserving $H(Z|X,Y,U,V)$ and the marginal on $(Z,V)$ retains the Markov chain $(X,Y,U)-V-Z$.  The resulting distribution is formed by an average over distributions of $(X,Y,Z,V)$ that satisfy $X-V-Y$, thus preserving the Markov chain $X-(U,V)-Y$ as well.

Note, that we can redefine $U:=(U,V)$ without loss of generality, so have the functional dependence $H(V|U)=0$.
Using  \eqref{vi.2} we only have that\vspace{-.1cm}
\begin{align}
\normtv{\Gamma_{XYZ}-Q_XQ_{YZ|X}}
&=\normtv{P_{X_TY_TZ_T}-Q_XQ_{YZ|X}}\\
&\le \normtv{P_{X^nY^nZ^n}-\prod Q_XQ_{YZ|X}}\\
&<\e.
\end{align}
So far we have shown that the rates for any feasible scheme lie in
\begin{equation}\s_{D_\e,\e}\defeq \setbr{\begin{array}{r c l}
\hspace{-.1cm}(R_0,R_1,R_2)\in\R^3&:&\ex P_{X,Y,Z,U,V}\in D_\e \mbox{ s.t.}\\
R_1&\ge&I(X;U,V)\\
R_2&\ge&I(X;V)\\
R_0&\ge&I(X,Y,Z;U,V)-\ldots\\
&&\quad(g_1(\e)+2g(\e))
\end{array}},\end{equation}where
\begin{equation}D_\e\defeq\setbr{\begin{array}{r c l}
P_{X,Y,Z,U,V}&:&\normtv{P_{XYZ}-Q_XQ_{YZ|X}}\le\e,\\
&&X-(U,V)-Y,\\
&&(X,Y,U)-V-Z,\\
&&\abs{\V}\le\abs{\X}\abs{\Y}\abs{\Z}+3,\\
&&\abs{\U}\le\abs{\X}\abs{\Y}\abs{\Z}\abs{\V}+2
\end{array}}.
\end{equation}

\subsection{Continuity of $\s_{D_\e,\e}$ at $\e=0$}

The final step crucially depends on the compactness endowed by the above cardinality bounds.  We would like to show that taking the limit $\e\downarrow0$ recovers $\s_D$ i.e.
\begin{equation}
\bigcap_{\e>0}\s_{D_\e,\e}=\s_D.
\end{equation}

First, note that
\begin{equation}
\bigcap_{\e>0}\s_{D_\e,\e}\supseteq\s_D
\end{equation}
since $\s_{D_\e,\e}$ shrinks as $\e$ shrinks, and $\s_{D_0,0}=\s_D$.

For the other direction, consider the stricter set
\begin{equation}\s'_{D_\e,\e}\defeq \setbr{\begin{array}{r c l}
\hspace{-.1cm}(R_0^2)\in\R^3&:&\ex P_{X,Y,Z,U,V}\in D_\e \mbox{ s.t.}\\
R_1&\ge&I(X;U,V)\\
R_2&\ge&I(X;V)\\
R_0&\ge&I(X,Y,Z;U,V)
\end{array}}.\end{equation}

Note that
\begin{equation}
\bigcap_{\e>0}\s_{D_\e,\e}\subseteq \clos\br{\bigcap_{\e>0}\s'_{D_\e,\e}}.\label{inc}
\end{equation}

Suppose otherwise: there exists a rate triple $(a,b,c)$ in the left-hand side (LHS), but not the right-hand side (RHS).  Then let's say $a^*$ is the smallest value of $R_0$ such that $(a^*,b,c)$ is in the RHS, so $a^*>a$ necessarily.  Since $a^*$ is the smallest value that guarantees inclusion in the RHS, we can pick $\e$ small enough to exclude $((a+a^*)/2,b,c)$ from $\s'_{D_\e,\e}$.  Also, we can pick $\e$ small enough to ensure that $g_1(\e)+2g(\e)<(a^*-a)/2$.  Both conditions together imply that for $\e$ small enough, we have
\begin{align}
(a+a^*)/2&<\min_{P_{X,Y,Z,U,V}\in D_\e}I(X,Y,Z;U,V)=:R_0^*\\
\iff a&<R_0^*+(a-a^*)/2\\
&<R_0^*-(g_1(\e)+2g(\e)),
\end{align}
i.e. $(a,b,c)$ is not in the LHS, which contradicts our initial assumption.  Thus, \eqref{inc} holds.

Next, consider the mapping
\begin{equation}
f(P_{XYZUV})=(I(X,Y,Z;U,V),I(X;U,V),I(X;V)),
\end{equation}
which yields the frontier of Pareto-optimal rates in the guise of $f(D)$ and $f(D_\e)$.  Note that
\begin{equation}
\bigcap_{\e>0}f(D_\e)=f(D),
\end{equation}
because $\bigcap_{\e>0}D_\e=D$, the sets $D_\e$ are decreasing subsets (as $\e$ decreases) of the compact probability simplex, and $f$ is a continuous function.  Thus, we have
\begin{equation}
\bigcap_{\e>0}\s'_{D_\e,\e}=\s_D.
\end{equation}
Finally, note that $\s_D$ is closed since $f$ is continuous and $D$ is compact.  This completes the converse proof.

\section{Summary}

Coordination in general networks remains a challenging area of study, as far as tight results are concerned.  In this work, we provide a tight result for strong coordination - generating correlated random variables - on a cascade network.  Even though the result extends to arbitrarily long networks, it is made tractable by a secrecy constraint that ensures that the synthesized sequences are independent of the communication transmissions.  We also demonstrated how requiring secrecy on a subset of the communication links expands the optimal rate region.  On the other hand, lack of a centralized source of common randomness was shown to shrink the optimal rate region and, in particular, the optimal communication rate region.  The questions of whether Conjecture 1 of \cite{coord} is true with a central source of common randomness, and whether networks can contribute to our understanding of the relationship between empirical coordination and strong coordination, form an exciting direction for future study.

\section*{Acknowledgement}

We would like to thank an anonymous reviewer for suggesting worthwhile additions to Sections III-A and III-B.


\section*{Appendix}

\emph{1) Proof for Arbitrarily Long Cascades (II-F):}
When we have $m\ge 3$ nodes in the cascade, the proof  follows the same steps outlined in sections IV and V.  

\emph{Achievability:} The idealized distribution $\Up$ used for the achievability proof is constructed as follows. For $n\ge1$, let $(K,M'_1,M'_2,\ldots,M'_{m-1})$ be uniformly distributed on $[2^{nR_0}]\times[2^{n(R_1-\sum_{j=2}^{m-1}R_j)}]\times\cdots\times[2^{nR_{m-1}}]$.  We shall set $M_1:=(M'_1,\ldots,M'_{m-1}), M_2:=(M'_2,\ldots,M'_{m-1})$ and so on till $M_{m-1}:=M'_{m-1}$.  Next, we use a superposition codebook with $(m-1)$ layers.

Consider a codebook $\B_{U_{m-1}^n}$ of $2^{n(R_0+R_{m-1})}$ $U_{m-1}^n$ sequences (randomly drawn according to the i.i.d.\ distribution $\prod Q_{U_{m-1}}$), where the codewords are indexed as $u_{m-1}^n(M'_{m-1},K)$.  For every $U_{m-1}^n$ codeword, we have a codebook $\B_{U_{m-2}^n|U_{m-1}^n}$ of $2^{n(R_{m-2}-R_{m-1})}$ $U_{m-2}^n$ sequences (randomly drawn by passing the $U_{m-1}^n$ codeword through the memoryless channel $\prod Q_{U_{m-2}|U_{m-1}}$), indexed as $u_{m-2}^n(M'_{m-2},M'_{m-1},K)$.  The following layers of the codebook are built in a recursive fashion, conditioned on all codewords that have been selected in lower layers.  The operational distribution $P$ is defined in analogous fashion to \eqref{scheme}, with
\begin{equation}
P_{Y_1^n,\ldots,Y_{m-1}^n,M'_1,\ldots,M'_{m-1}|X^n,K}=\Up_{Y_1^n,\ldots,Y_{m-1}^n,M'_1,\ldots,M'_{m-1}|X^n,K}.
\end{equation}

The total variation analysis, to show the existence of SCCS codes, follows section IV-D.  The analysis of the communication rates requires the general version of Lemma \ref{gensoft} found in \cite{Gohari1}.

\emph{Converse:} The steps here parallel those in section V-B.  Consider a secure channel synthesis code that satisfies
\begin{equation}
\normtv{P_{XY^{m-1}M^{m-1}}-P_{M^{m-1}}\prod Q_{XY^{m-1}}}<\e,\label{genach}
\end{equation}
for $\e\in(0,1/4)$ and $n$ sufficiently large.
For the $i$th communication rate ($1\le i\le m-1$), we have
\begin{align}
nR_i&\ge H(M_i)\\
&\ge H(M_i|K)\\
&\ge I(X^n;M_i|K)\\
&= I(X^n;M_i,M_{i+1},\ldots,M_{m-1}|K)\\
&=I(X^n;M_i^{m-1},K)\\
&=\sum_{i=1}^nI(X_i;M_i^{m-1},K|X^{i-1})\\
&=\sum_{i=1}^nI(X_i;M_i^{m-1},K,X^{i-1})\\
&\ge\sum_{i=1}^nI(X_i;M_i^{m-1},K)\\
&=nI(X_T;M_i^{m-1},K|T)\\
&=nI(X_T;M_i^{m-1},K,T).
\end{align}
For the common randomness rate, we have
\begin{align}
nR_0
&\ge H(K)\ge H(K|M^{m-1})\\
&\ge I(X^n,Y_1^n,\ldots,Y_{m-1}^n;K|M^{m-1})\\
&\ge I(X^n,Y_1^n,\ldots,Y_{m-1}^n;M^{m-1},K)-n f_1(\e)\\
&\ge\sum_{t=1}^nI(X_t,(Y_1)_t,\ldots,(Y_{m-1})_t;M^{m-1},K)-nf_1(\e)-nf_2(\e)\\
&\ge nI(X_T,(Y_1)_T,\ldots,(Y_{m-1})_T;M^{m-1},K|T)-n(f_1(\e)+f_2(\e))\\
&\ge nI(X_T,(Y_1)_T,\ldots,(Y_{m-1})_T;M^{m-1},K,T)-n(f_1(\e)+2f_2(\e)),
\end{align}
where the approximate inequalities follow from \eqref{genach}, with $\lim_{\e\downarrow0}f_1(\e)=\lim_{\e\downarrow0}f_2(\e)=0$.  The cardinality bound \eqref{gencard} is derived by using the Carath\'{e}odory theorem, as demonstrated in section V-B.  Finally, the converse proof is completed by letting $\e\downarrow 0$ and invoking compactness of the set of distributions that define the optimal rate-region, as done in section V-C.\\

\emph{2) Proof for Theorem 2 (No Secrecy on Link 1):}\\
\emph{Achievability:} We rely on the same idealized distribution $\Up$ used in the proof of Theorem 1, as depicted in Fig.\ \ref{ideal}.  The operational distribution $P$ is defined in \eqref{scheme}. Note that by the triangle inequality and \eqref{v.1}, we have (expectation is over the $(U^n,V^n)$ codebook)
\begin{align}
\E\normtv{P_{X^nY^nZ^nM_2}-P_{M_2}\prod Q_{XYZ}}
&\le \E\normtv{P_{X^nY^nZ^nM_2}-\Up_{X^nY^nZ^nM_2}}+\cdots\nonumber\\
&\quad~~\E\normtv{\Up_{X^nY^nZ^nM_2}-\Up_{M_2}\prod Q_{XYZ}}+\cdots\nonumber\\
&\quad~~\E\normtv{\Up_{M_2}\prod Q_{XYZ}-P_{M_2}\prod Q_{XYZ}}\\
&= \E\normtv{P_{X^nY^nZ^nM_2}-\Up_{X^nY^nZ^nM_2}}+\cdots\nonumber\\
&\quad~~\E\normtv{\Up_{X^nY^nZ^nM_2}-\Up_{M_2}\prod Q_{XYZ}}+\cdots\nonumber\\
&\quad~~\E\normtv{\Up_{M_2}-P_{M_2}}\\
&\le 2\E\normtv{P_{X^nY^nZ^nM_2}-\Up_{X^nY^nZ^nM_2}}+\cdots\nonumber\\
&\quad~~\E\normtv{\Up_{X^nY^nZ^nM_2}-\Up_{M_2}\prod Q_{XYZ}}.\label{interv2}
\end{align}
Again, we have reduced the problem to showing that
\begin{itemize}
\item the idealized distribution $\Up$ satisfies \eqref{locsec}, and
\item the operational distribution $P$ is well-approximated by $\Up$.
\end{itemize}

The latter can be shown in identical fashion to the steps in section IV-D, after observing that
\begin{equation}
\E\normtv{P_{X^nY^nZ^nM_2}-\Up_{X^nY^nZ^nM_2}}\le \E\normtv{P_{X^nK}-\Up_{X^nK}}.
\end{equation}
The resulting communication rate requirements are
\begin{align}
R_2&> I(X;V),\label{r2v2}\\
R_1&> I(X;U,V).
\end{align}

It remains to bound the second term in \eqref{interv2}. Note that with fixed $m_b$, the scheme may choose from $2^{n(R_0+R_1-R_2)}$ $W^n$ codewords, where $W\defeq (U,V)$. Since the codebook has a superposition structure, by arguments similar to those made in the analysis of the communication rates, we have for any $m_b$ that
\begin{equation}
\E\normtv{\Up_{X^nY^nZ^n|M_2=m_b}-\prod Q_{XYZ}}<\e_n,
\end{equation}
with $\lim_{n\to\infty}\e_n=0$, when
\begin{align}
R_0&> I(X,Y,Z;V),\\
(R_1-R_2)+R_0&> I(X,Y,Z;U,V)\\
\Rightarrow R_1+R_0&> I(X,Y,Z;U,V)+I(X;V),
\end{align}
where the final inequality follows from \eqref{r2v2} .  This implies that
\begin{align}
\E\normtv{\Up_{X^nY^nZ^nM_2}-\Up_{M_2}\prod Q_{XYZ}}
&=\E\frac{1}{2}\sum_{x^ny^nz^nm_2}\abs{\Up_{X^nY^nZ^nM_2}-\frac{1}{2^{nR_1}}\prod Q_{XYZ}}\\
&=\frac{1}{2^{nR_1}}\E\frac{1}{2}\sum_{x^ny^nz^nm_2}\abs{\Up_{X^nY^nZ^n|M_2}-\prod Q_{XYZ}}\\
&=\frac{1}{2^{nR_1}}\sum_{m_2}\E\normtv{\Up_{X^nY^nZ^n|M_2}-\prod Q_{XYZ}}\\
&<\frac{1}{2^{nR_1}}\sum_{m_2}\e_n\\
&=\e_n,\label{ach1v2}
\end{align}
so indeed, $\Up$ satisfies \eqref{locsec}.

\emph{Converse:} The steps here parallel those in section V-B.  Consider a secure channel synthesis code that satisfies
\begin{equation}
\normtv{P_{X^nY^nZ^nM_2}-P_{M_2}\prod Q_{XYZ}}<\e,\label{locach}
\end{equation}
for $\e\in(0,1/4)$ and $n$ sufficiently large.
Using the steps in section IV-B, we bound the communication rates as
\begin{align}
R_1&\ge I(X_T;M_1,M_2,K,T),\\
R_2&\ge I(X_T;M_2,K,T).
\end{align}
For the common randomness rate, we have
\begin{align}
nR_0
&\ge H(K)\ge H(K|M_2)\\
&\ge I(X^n,Y^n,Z^n;K|M_2)\\
&\ge I(X^n,Y^n,Z^n;M_2,K)-ng_1(\e)\\
&\ge\sum_{t=1}^nI(X_t,Y_t,Z_t;M_2,K)-ng_1(\e)-ng(\e)\\
&\ge nI(X_T,Y_T,Z_T;M_2,K|T)-n(g_1(\e)+g(\e))\\
&\ge nI(X_T,Y_T,Z_T;M_2,K,T)-n(g_1(\e)+2g(\e)),
\end{align}
where the approximate inequalities follow from \eqref{locach}, with $\lim_{\e\downarrow0}g(\e)=\lim_{\e\downarrow0}g_1(\e)=0$.  
Similarly, we have the sum rate constraint
\begin{align}
n(R_1+R_0)
&\ge H(M_1,K)\\
&\ge H(M_1,K|M_2)\\
&\ge I(X^n,Y^n,Z^n;M_1,K|M_2)\\
&\ge I(X^n,Y^n,Z^n;M_1,M_2,K)-ng_1(\e)\\
&\ge\sum_{t=1}^nI(X_t,Y_t,Z_t;M_1,M_2,K)-ng_1(\e)-ng(\e)\\
&\ge nI(X_T,Y_T,Z_T;M_1,M_2,K|T)-n(g_1(\e)+g(\e))\\
&\ge nI(X_T,Y_T,Z_T;M_1,M_2,K,T)-n(g_1(\e)+2g(\e)).
\end{align}

Setting $X=X_T,Y=Y_T,Z=Z_T,U=(M_1)$ and $V=(M_2,K,T)$ results in the rate expressions provided in \eqref{locsecout} with the Markov chains in \eqref{D}.
The cardinalities of $U$ and $V$ are bounded by using the Carath\'{e}odory theorem, as demonstrated in section V-B.  Finally, the converse proof is completed by letting $\e\downarrow 0$ and invoking compactness of the set of distributions that define the optimal rate-region, as done in section V-C.\\

\emph{Task Assignment Example:}
Since
\begin{align}
I(X,Y,Z;U,V)+I(X;V)&=I(X,Y,Z;V)+I(X,Y,Z;U|V)+I(X;V)\\
&=I(X,Y,Z;V)+I(X;U,V)+I(Y,Z;U|V,X)\\
&=I(X,Y,Z;V)+I(X;U,V)+I(Y;U|V,X),
\end{align}
the result is tight for any distribution which can be efficiently synthesized under the constraint $I(Y;U|V,X)=0$.
The sum-rate constraints in \eqref{locsecin} and \eqref{locsecout} are redundant in this case.

When $(X,Y,Z)$ is a random permutation of $\setbr{1,2,3}$, then $(X,Y,U)-V-Z\Rightarrow H(Z|V)=0$, since $Z=\setbr{1,2,3}\setminus\setbr{X,Y}$.  We also have that $Y=\setbr{1,2,3}\setminus\setbr{X,Z}$, so
\begin{align}
I(Y;U|V,X)&=I(Y;U|V,Z,X)\\
&=0,
\end{align}
since $Y$ is a function of $(X,Z)$.  Thus, the sum-rate constraints in \eqref{locsecin} and \eqref{locsecout} become redundant, and \eqref{locsecin} reduces to \eqref{locsecout}.\\

\emph{3) Proof for Theorem 3 (No Secrecy on Link 2):}\\
\emph{Achievability:} Using \eqref{v.1}, we have
\begin{equation}
\normtv{P_{X^nY^nZ^nM_1}-P_{M_1}\prod Q_{XYZ}} \le \normtv{P_{X^nY^nZ^nM_1M_2}-P_{M_1M_2}\prod Q_{XYZ}},
\end{equation}
so the achievability part readily follows from the achievability proof of Theorem 1.  Note that $\s_{D'}^{(loc)}$ is the region $\s_{D'}$ in Theorem 1.

\emph{Converse:} The steps here parallel those in section V-B.  Consider a secure channel synthesis code that satisfies
\begin{equation}
\normtv{P_{X^nY^nZ^nM_1}-P_{M_1}\prod Q_{XYZ}}<\e,\label{locach1}
\end{equation}
for $\e\in(0,1/4)$ and $n$ sufficiently large.
Using the steps in section IV-B, we bound the communication rates as
\begin{align}
R_1&\ge I(X_T;M_1,K,T),\\
R_2&\ge I(X_T;M_2,K,T).
\end{align}
For the common randomness rate, we have
\begin{align}
nR_0
&\ge H(K)\ge H(K|M_1)\\
&\ge I(X^n,Y^n,Z^n;K|M_1)\\
&\ge I(X^n,Y^n,Z^n;M_1,K)-ng_1(\e)\\
&\ge\sum_{t=1}^nI(X_t,Y_t,Z_t;M_1,K)-ng_1(\e)-ng(\e)\\
&\ge nI(X_T,Y_T,Z_T;M_1,K|T)-n(g_1(\e)+g(\e))\\
&\ge nI(X_T,Y_T,Z_T;M_1,K,T)-n(g_1(\e)+2g(\e)),
\end{align}
where the approximate inequalities follow from \eqref{locach1}, with $\lim_{\e\downarrow0}g(\e)=\lim_{\e\downarrow0}g_1(\e)=0$.  

Setting $X=X_T,Y=Y_T,Z=Z_T,U=(M_1,K,T)$ and $V=(M_2,K,T)$ results in the rate expressions provided in \eqref{locsec1reg} with the Markov chains in \eqref{D}.
The cardinalities of $U$ and $V$ are bounded by using the Carath\'{e}odory theorem, as demonstrated in section V-B.  Finally, the converse proof is completed by letting $\e\downarrow 0$ and invoking compactness of the set of distributions that define the optimal rate-region, as done in section V-C.\\

\emph{4) Proof for Theorem 4 (Cascade with Relay):}\\
\emph{Achievability:} We stitch together two point-to-point strong coordination schemes in order to achieve \eqref{relay}.  From \cite[Theorem II.1]{DCS}, we know that $R_1>I(X;U)$ suffices to achieve
\begin{equation}
\normtv{P_{X^nU^n}-\prod Q_{XU}}<\e
\end{equation}
for any $\e>0$ and $n$ sufficiently large. The relay node generates a $U^n$ sequence and then synthesizes a channel to $Z^n$ at rate $R_2>I(Z;U)$, so we have
\begin{equation}
\normtv{P_{U^nZ^n}-\prod Q_{UZ}}<\e
\end{equation}
for any $\e>0$ and $n$ sufficiently large.  Note that we have $X^n-U^n-Z^n$ under the operational distribution $P$.  Also, the above constraints imply that
\begin{equation}
\normtv{P_{U^n}-\prod Q_{U}}<\e
\end{equation}
for $n$ sufficiently large, by \eqref{v.1}. Now, it remains to argue that $(X^n,Z^n)$ is approximately i.i.d.  The following steps make a statement that is analogous to the Markov lemma \cite{berger}, commonly used to prove results about empirical coordination.  We use the triangle inequality, \eqref{v.1} and \eqref{v.2} below, along with the above statements.  We have
\begin{align}
\normtv{P_{X^nZ^n}-\prod Q_{XZ}}&=\normtv{\sum_{u^n}P_{U^n}P_{X^n|U^n}P_{Z^n|U^n}-\sum_{u^n}\prod Q_UQ_{X|U}Q_{Z|U}}\\
&\le \normtv{\sum_{u^n}\br{\prod Q_{UZ}}P_{X^n|U^n}-\sum_{u^n}\br{\prod Q_{UZ}}\prod Q_{X|U}}+\cdots\nonumber\\
&\qquad \normtv{\sum_{u^n}P_{U^nZ^n}P_{X^n|U^n}-\sum_{u^n}\br{\prod Q_{UZ}}P_{X^n|U^n}}\\
&\le \normtv{\br{\prod Q_{UZ}}P_{X^n|U^n}-\br{\prod Q_{UZ}}\prod Q_{X|U}}+\cdots\nonumber\\
&\qquad \normtv{P_{U^nZ^n}P_{X^n|U^n}-\br{\prod Q_{UZ}}P_{X^n|U^n}}\\
&=\frac{1}{2}\sum_{x^n,u^n,z^n} \br{\prod Q_{UZ}}\abs{P_{X^n|U^n}-\prod Q_{X|U}}+\cdots\nonumber\\
&\qquad \normtv{P_{U^nZ^n}-\br{\prod Q_{UZ}}}\\
&<\normtv{\br{\prod Q_{U}}P_{X^n|U^n}-\prod Q_{XU}}+\e\\
&\le \normtv{P_{X^nU^n}-\prod Q_{XU}}+\normtv{\br{\prod Q_{U}}P_{X^n|U^n}-P_{U^n}P_{X^n|U^n}}+\e\\
&<\normtv{\br{\prod Q_{U}}-P_{U^n}}+2\e\\
&<3\e,
\end{align}
for $n$ sufficiently large.\\

\emph{Converse:} We assume that
\begin{equation}
\normtv{P_{X^nZ^n}-\prod Q_{XZ}}<\e\label{relayach}
\end{equation}
for $\e\in(0,1/4)$ and $n$ sufficiently large.

Since $X^n$ is i.i.d., we have
\begin{align}
nR_1&\ge H(M_1)\\
&\ge I(X^n;M_1|K_1)\\
&=I(X^n;M_1,K_1)\\
&=\sum_{i=1}^nI(X_i;M_1,K_1|X^{i-1})\\
&=\sum_{i=1}^nI(X_i;M_1,K_1,X^{i-1})\\
&\ge \sum_{i=1}^nI(X_i;M_1,K_1)\\
&=nI(X_T;M_1,K_1|T)\\
&=nI(X_T;M_1,K_1,T),
\end{align}
where $T$ is a random time index uniformly distributed on $[n]$.  To bound $R_2$, we make use of the fact that $K_2\perp (M_1,K_1)$, since $M_1-(X^n,K_1)-K_2$, and $K_2 \perp (X^n,K_1)$.  Also, we use the constraint that $(M_1,K_1)-(M_2,K_2)-Z^n$. Consider
\begin{align}
nR_2&\ge H(M_2)\\
&\ge I(Z^n;M_2|K_2)\\
&= I(Z^n;M_2,K_2)-I(Z^n;K_2)\\
&=I(Z^n;M_1,K_1,M_2,K_2)-I(Z^n;K_2)\\
&=I(Z^n;M_1,K_1)+I(Z^n;M_2,K_2|M_1,K_1)-I(Z^n;K_2)\\
&=I(Z^n;M_1,K_1)+I(Z^n;K_2|M_1,K_1)+I(Z^n;M_2|M_1,K_1,K_2)-I(Z^n;K_2)\\
&=I(Z^n;M_1,K_1)+I(Z^n,M_1,K_1;K_2)+I(Z^n;M_2|M_1,K_1,K_2)-I(Z^n;K_2)\\
&=I(Z^n;M_1,K_1)+I(M_1,K_1;K_2|Z^n)+I(Z^n;M_2|M_1,K_1,K_2)\\
&\ge I(Z^n;M_1,K_1)\\
&=\sum_{i=1}^nI(Z_i;M_1,K_1|Z^{i-1})\\
&\ge\sum_{i=1}^nI(Z_i;M_1,K_1)-g_2(\e)\\
&=nI(Z_T;M_1,K_1|T)-g_2(\e)\\
&\ge nI(Z_T;M_1,K_1,T)-2g_2(\e),\label{r2app}
\end{align}
where the approximate inequalities follow from \eqref{relayach}, with $\lim_{\e\downarrow0}g_2(\e)=0$.

Setting $X=X_T,Z=Z_T,U=(M_1,K_1,T)$ results in the rate expressions provided in \eqref{relayresult} with $X-U-Z$, up to the correction in \eqref{r2app}.
The cardinality of $U$ is bounded by using the Carath\'{e}odory theorem, as demonstrated in section V-B.  Finally, the converse proof is completed by letting $\e\downarrow 0$ and invoking compactness of the set of distributions that define the optimal rate-region, as done in section V-C.\\



\bibliographystyle{ieeetr}

\bibliography{sccs}

\begin{thebibliography}{10}

\bibitem{sai}
S.~Satpathy and P.~Cuff, ``Secure cascade channel synthesis,'' in {\em
  Information Theory Proceedings (ISIT), 2013 IEEE International Symposium on},
  pp.~2955--2959, July 2013.

\bibitem{coord}
P.~Cuff, H.~H. Permuter, and T.~M. Cover, ``Coordination capacity,'' {\em IEEE
  Trans. on Info. Theory}, vol.~56, no.~9, pp.~4181--4206, 2010.

\bibitem{DCS}
P.~Cuff, ``Distributed channel synthesis,'' {\em Information Theory, IEEE
  Transactions on}, vol.~59, pp.~7071--7096, Nov 2013.

\bibitem{Cuff1}
P.~Cuff, ``Communication requirements for generating correlated random
  variables,'' in {\em ISIT}, pp.~1393--1397, 2008.

\bibitem{reverse2}
C.~Bennett, I.~Devetak, A.~Harrow, P.~Shor, and A.~Winter, ``The quantum
  reverse shannon theorem and resource tradeoffs for simulating quantum
  channels,'' {\em Information Theory, IEEE Transactions on}, vol.~60,
  pp.~2926--2959, May 2014.

\bibitem{Gohari1}
A.~Gohari and V.~Anantharam, ``Generating dependent random variables over
  networks,'' in {\em ITW}, pp.~698 --702, Oct. 2011.

\bibitem{Gohari2}
A.~Gohari, M.~Yassaee, and M.~Aref, ``Secure channel simulation,'' in {\em
  Information Theory Workshop (ITW), 2012 IEEE}, pp.~406--410, Sept 2012.

\bibitem{Gohari3}
M.~Yassaee, A.~Gohari, and M.~Aref, ``Channel simulation via interactive
  communications,'' {\em Information Theory, IEEE Transactions on}, vol.~61,
  pp.~2964--2982, June 2015.

\bibitem{Gohari4}
F.~Haddadpour, M.~H. Yassaee, A.~Gohari, and M.~R. Aref, ``Coordination via a
  relay,'' in {\em ISIT}, pp.~3048--3052, 2012.

\bibitem{Wyner}
A.~Wyner, ``The common information of two dependent random variables,'' {\em
  IEEE Transactions on Information Theory}, vol.~21, pp.~163--179, Mar 1975.

\bibitem{Haim}
T.~M. Cover and H.~H. Permuter, ``Capacity of coordinated actions,'' in {\em
  ISIT}, pp.~2701 --2705, June 2007.

\bibitem{Cover}
T.~M. Cover and J.~A. Thomas, {\em Elements of information theory (2. ed.)}.
\newblock Wiley, 2006.

\bibitem{Cuff2}
P.~Cuff, ``State information in bayesian games,'' in {\em Allerton}, 2009.

\bibitem{RDSS}
C.~Schieler and P.~Cuff, ``Rate-distortion theory for secrecy systems,'' {\em
  Information Theory, IEEE Transactions on}, vol.~60, pp.~7584--7605, Dec 2014.

\bibitem{Winter05}
A.~Winter, ``Secret, public and quantum correlation cost of triples of random
  variables,'' in {\em ISIT 2005}, pp.~2270--2274, 2005.

\bibitem{Winter14}
E.~Chitambar, M.~H. Hsieh, and A.~Winter, ``The private and public correlation
  cost of three random variables with collaboration,'' {\em IEEE Transactions
  on Information Theory}, vol.~62, pp.~2034--2043, April 2016.

\bibitem{Bloch2}
M.~Bloch and J.~Laneman, ``Strong secrecy from channel resolvability,'' {\em
  Information Theory, IEEE Transactions on}, vol.~59, pp.~8077--8098, Dec 2013.

\bibitem{control}
R.~Murray, K.~Astrom, S.~Boyd, R.~Brockett, and G.~Stein, ``Future directions
  in control in an information-rich world,'' {\em Control Systems, IEEE},
  vol.~23, pp.~20 -- 33, apr 2003.

\bibitem{casca1}
H.~Yamamoto, ``Source coding theory for cascade and branching communication
  systems,'' {\em Information Theory, IEEE Transactions on}, vol.~27,
  pp.~299--308, May 1981.

\bibitem{casca2}
D.~Vasudevan, C.~Tian, and S.~N. Diggavi, ``Lossy source coding for a cascade
  communication system with side-information,'' in {\em Allerton}, 2006.

\bibitem{casca4}
H.~Permuter and T.~Weissman, ``Cascade and triangular source coding with side
  information at the first two nodes,'' {\em Information Theory, IEEE
  Transactions on}, vol.~58, pp.~3339--3349, June 2012.

\bibitem{casca3}
P.~Cuff, H.-I. Su, and A.~El~Gamal, ``Cascade multiterminal source coding,'' in
  {\em Information Theory, 2009. ISIT 2009. IEEE International Symposium on},
  pp.~1199--1203, June 2009.

\bibitem{Bloch3}
M.~Bloch and J.~Kliewer, ``Strong coordination over a line network,'' in {\em
  Information Theory Proceedings (ISIT), 2013 IEEE International Symposium on},
  pp.~2319--2323, July 2013.

\bibitem{Vell}
B.~N. Vellambi, J.~Kliewer, and M.~R. Bloch, ``Strong coordination over
  multi-hop line networks,'' in {\em Information Theory Workshop - Fall (ITW),
  2015 IEEE}, pp.~192--196, Oct 2015.

\bibitem{Vell1}
B.~N. Vellambi, J.~Kliewer, and M.~R. Bloch, ``Strong coordination over
  multi-hop line networks,'' {\em CoRR}, vol.~abs/1602.09001, 2016.

\bibitem{Cuff6}
P.~Cuff, ``Secrecy in cascade networks,'' in {\em Information Theory Workshop
  (ITW), 2013 IEEE}, pp.~1--5, Sept 2013.

\bibitem{comm}
W.~Liu, G.~Xu, and B.~Chen, ``The common information of n dependent random
  variables,'' in {\em Communication, Control, and Computing (Allerton), 2010
  48th Annual Allerton Conference on}, pp.~836--843, Sept 2010.

\bibitem{resolvability}
T.~Han and S.~Verd\'{u}, ``Approximation theory of output statistics,'' {\em
  Info. Theory, IEEE Trans. on}, vol.~39, pp.~752 --772, may 1993.

\bibitem{eva}
E.~Song, P.~Cuff, and H.~Poor, ``The likelihood encoder for lossy source
  compression,'' in {\em Information Theory (ISIT), 2014 IEEE International
  Symposium on}, pp.~2042--2046, June 2014.

\bibitem{Gohari5}
M.~Yassaee, M.~Aref, and A.~Gohari, ``Achievability proof via output statistics
  of random binning,'' {\em Information Theory, IEEE Transactions on}, vol.~60,
  pp.~6760--6786, Nov 2014.

\bibitem{Bloch1}
M.~Bloch, L.~Luzzi, and J.~Kliewer, ``Strong coordination with polar codes,''
  in {\em Communication, Control, and Computing (Allerton), 2012 50th Annual
  Allerton Conference on}, pp.~565--571, Oct 2012.

\bibitem{egg}
H.~G. Eggleston, J.~A. Todd, and F.~Smithies, ``Convexity,'' 1966.

\bibitem{berger}
T.~Berger, ``Multiterminal source coding,'' {\em The information theory
  approach to communications}, vol.~229, pp.~171--231, 1977.

\end{thebibliography}

\end{document}